\documentclass[12pt,letterpaper]{article}
\usepackage{epsfig,amsmath}
\usepackage{amssymb,amsmath,wasysym}
\usepackage{bbm}	
\usepackage[normalem]{ulem} 

\newcommand{\beq}{\begin{equation}}
\newcommand{\eeq}{\end{equation}}
\newcommand{\beqarr}{\begin{eqnarray}}
\newcommand{\eeqarr}{\end{eqnarray}}

\title{Perturbation Analysis of the Kuramoto Phase Diffusion Equation Subject to Quenched Frequency Disorder}
\author{Ralf T\"onjes, Bernd Blasius}

\begin{document}

\maketitle

\begin{abstract}
\noindent
The Kuramoto phase diffusion equation is a nonlinear partial differential equation which describes the spatio-temporal evolution of a phase variable in an oscillatory reaction diffusion system. 
Synchronization manifests itself in a stationary phase gradient where all phases throughout a system evolve with the same velocity, the synchronization frequency. The formation of concentric waves can be explained by local impurities of higher frequency which can entrain their surroundings. Concentric waves in synchronization also occur in heterogeneous systems, where the local frequencies are distributed randomly.
We present a perturbation analysis of the synchronization frequency where the perturbation is given by the heterogeneity of natural frequencies in the system. The nonlinearity in form of dispersion, leads to an overall acceleration of the oscillation for which the expected value can be calculated from the second order perturbation terms. We apply the theory to simple topologies, like a line or the sphere, and deduce the dependence of the synchronization frequency on the size and the dimension of the oscillatory medium. We show that our theory can be extended to include rotating waves in a medium with periodic boundary conditions.
By changing a system parameter the synchronized state may become quasi degenerate. We demonstrate how perturbation theory fails at such a critical point.
\end{abstract}

\noindent
\section*{I. Introduction}

The formation of spatio-temporal patterns is ubiquitous in natural and artificial complex dynamical systems \cite{Kuramoto84,Mikhailov90,Kapral95}. In oscillatory media pattern formation    is tightly connected to the process of synchronization and 
plays an important role in a variety of systems far from equilibrium, such as arrays of Josephson junctions \cite{Wiesenfeld96}, the Beluzov-Zhabotinsky reaction \cite{ZaiZhab70,ShowBla07}, cardiac tissue \cite{Winfree94}, neural systems \cite{Varela01} 
and spatially extended ecological systems \cite{BlaStone99,BlaStone00,Blasius07}. 
Different mechanisms for pattern formation are known. One of these is the interplay between attractive interaction, e.g. diffusion, which mediates long range correlations, and heterogeneity, or disorder, driving the system away from a uniform state. However, while large amounts of spatial heterogeneity describe the reality of most natural and biological systems, not much about the pattern formation and synchronization in disordered oscillatory media is known.
\\ \\
Synchronization in the sense of a mutual adjustment of internal frequencies
\cite{PiRoKurths03} does not necessarily imply a total reduction of the system dimension to that of a single component, i.e.~completely uniform dynamics. Instead, even in the synchronized state parameters like phase can vary across the system while the phase differences remain bounded or locked. In that case one can define waves that travel along a phase gradient \cite{Winfree01}. 
If the wavelength is smaller than the diameter of the system these waves are perceived as time periodic spatial patterns. Such waves are a prominent feature in regular low-dimensional reactor topologies of chemical oscillating reaction-diffusion systems \cite{ShowBla07}. Wave propagation in oscillatory systems, although experimentally more difficult to assess, is also observed and of much relevance in a biological, medical, ecological and epidemiological context \cite{Winfree94,BlaStone99,BlaStone00,JohnBjoeLieb04}. 
\\ \\
Beside spiral waves and turbulence, concentric ring waves patterns are one of the most prominent features in two dimensional oscillatory media. They are usually associated with the presence of local impurities in the system \cite{Kuramoto84,ZaiZhab70}. These pacemakers change the local oscillation frequency and are able to entrain their surroundings, which finally results in regular ring or target patterns.
However, concentric waves of surprising regularity occur also in heterogeneous systems, where the local frequencies are distributed randomly. This was first reported and explained in \cite{Sakaguchi88} and subsequently also observed in chaotic phase coherent systems \cite{BlaStone99,BlaStone00,Blasius07}.
In \cite{BlaToe05} it was shown that a phenomenological description can be obtained with phase equations using the Kuramoto model \cite{Kuramoto84}. The analysis revealed 
that the random nature of the medium itself plays a key role in the formation of the patterns. 
In order to utilize or control these patterns, in general, it will not be sufficient to understand the mechanisms leading to pattern formation. Of equally importance is a knowledge about the time and length scales involved  \cite{EngelSchoell06}. However, in disordered systems no analytic formula for such quantities, let alone the full phase profile, are known.
Here we show how estimations can be obtained by perturbation theory.
\\ \\
The objective of this paper is to derive first and second order perturbation terms for the synchronization frequency in the nonlinear Kuramoto Phase Diffusion equation (KPDE) given a time-independent distribution of frequencies in the system. We present two approaches to this problem. The first approach, as described in Section II, is based on a direct perturbation expansion of the KDPE. It includes as a special case, for vanishing nonlinearity, the exactly solvable situation of the inhomogeneous heat equation. The second approach presented in Section III is based on a Cole-Hopf transformation of the KPDE to a stationary Schr\"odinger equation for a particle in a disordered potential. Here, classical Schr\"odinger perturbation theory can be applied and leads to the same expressions as the first approach. We extend the second approach to describe variations in systems with topological charges. Such solutions exist for system topologies with periodic boundary conditions. 
In Section IV we apply the theory to simple topologies and calculate the first and second order perturbation terms of the synchronization frequency at the example of a $d$-dimensional medium with topological charges and also for the two dimensional surface of a sphere. 
Throughout we confirm our analytic results by direct numerical simulations. We deduce the dependence of the synchronization frequency on the size and the dimension of the oscillatory medium and demonstrate that the second order perturbation term changes its scaling behavior at the critical dimension $d=2$. 
Below that dimension it diverges with the system size and for $d>2$ it diverges for small frequency correlation lengths. Finally in Section V, we take a look at regimes which can not be described by perturbation theory. In particular, we observe a discontinuous change of the location of a dominant pacemaker center.
\\ \\
Let us start by reviewing the nonlinear phase diffusion equations \cite{Kuramoto84}. A heterogeneous oscillatory reaction diffusion system may be described by its full dynamics
\beq	\label{Eq:CompleteDyn}
	\dot{\mathbf{X}} = \mathbf{F}(\mathbf{X},\mathbf{r}) + \mathbf{\nabla}^2 \mathbf{D} \mathbf{X}(\mathbf{r})
\eeq
where $\mathbf{F}(\mathbf{X},\mathbf{r})$ describes an oscillatory nonequilibrium reaction at a position $\mathbf{r}$ given the vector of reactant concentrations $\mathbf{X}$ and the diffusion $\mathbf{\nabla}^2 \mathbf{D} \mathbf{X}(\mathbf{r})$ in the system, where $\mathbf{D}$ is a diagonal matrix of diffusion coefficients and the second order spatial derivative $\mathbf{\nabla}^2$ has to be applied component wise. 
Here we always assume that the local dynamics at the different locations are stable limit cycle oscillations.
If the diffusive coupling only leads to small deviations from these limit cycles and if the medium is locally isotropic the system can be reduced \cite{Mikhailov90} to the dynamics of phases $\vartheta(\mathbf{r})$ of the form
\beq	\label{Eq:KPDEs01}
	\dot{\vartheta}(\mathbf{r}) = \omega(\mathbf{r}) + \nabla^2\vartheta(\mathbf{r}) 
	                              + \gamma\left(\nabla\vartheta(\mathbf{r})\right)^2 \quad.
\eeq
These simplified phase equations define the nonlinear Kuramoto Phase Diffusion equations (KPDE). They were introduced by Kuramoto in 1976 \cite{Kuramoto76} and are obtained by the perturbative method of phase reduction, using averaging techniques, described in his seminal monograph from 1984 \cite{Kuramoto84}. 
Here, $\omega(\mathbf{r})$ is the local natural frequency of oscillation,
we have used a scaling of time to make the diffusion coefficient in front of the Laplacian differential operator equal to one and
the parameter $\gamma$ controls the nonlinearity, or dispersion.
It can directly be interpreted as the nonisochronicity, which is the shear rate of the phase flow near the limit cycle and describes the sensitivity of the phase velocities to changes in the oscillation amplitude \cite{Kuramoto84}.
\\ \\
The heterogeneity in the system may be parameterized by the sample variance
\beq	\label{Eq:sgmDef01}
	\sigma^2 =  \left\langle\omega^2\right\rangle_\textnormal{System} - \bigl\langle\omega\bigr\rangle^2_\textnormal{System}
\eeq
or some norm of the two point correlation function $C(\mathbf{r},\mathbf{r}')$ if the frequencies are random (but quenched), e.g.
\beq	\label{Eq:sgmDef02}
	\mathbb{E}\left[\omega(\mathbf{r})\omega(\mathbf{r}')\right] - \mathbb{E}\left[\omega^2\right]= \sigma^2 C(\mathbf{r},\mathbf{r}') \quad\textnormal{with}\quad ||C||=1 \,.
\eeq
For the phase equations (\ref{Eq:KPDEs01}) to be applicable to the problem Eq. (\ref{Eq:CompleteDyn}) the relaxation time of the amplitudes must be small compared to the time scale of the phase evolution \cite{Mikhailov90}, i.e. $\sigma\gamma\ll 1$. 
In the following we define $\omega=\sigma\eta$ with normalized frequencies $\eta$ 
and use $\sigma$ as a parameter of the system.
%
%
\\ \\
For a simulation of the KPDE (\ref{Eq:KPDEs01}) on a discretization of the medium it is of advantage to use the discrete Kuramoto model \cite{Kuramoto84}
\beq	\label{Eq:KPEs01}
	\dot\vartheta_n = \omega_n + \sum_{m} A_{nm} \sin(\vartheta_m-\vartheta_n) 
					    + B_{nm}~ \gamma~\bigl(1-\cos(\vartheta_m-\vartheta_n)\bigr)
\eeq
where the Laplacian of the medium is defined through the values $A_{nm}$ and the square absolute value of the gradient through the choices of $B_{nm}$. On a square lattice with nearest neighbor coupling of spacing $h$ we have $A_{nm}=B_{nm}=h^{-2}$.

\section*{II. Perturbation Approach 1}
In synchronization the phase velocities of Eq.~(\ref{Eq:KPDEs01})  have adapted to a common frequency
\beq	\label{Eq:KPDEs02}
\dot{\vartheta}(\mathbf{r}) ~=~ \Omega ~=~ \sigma\eta(\mathbf{r}) ~+~ \nabla^2\vartheta(\mathbf{r}) 
	                              ~+~ \gamma\left[\nabla\vartheta(\mathbf{r})\right]^2
\eeq
and with $\frac{d}{dt}\nabla\vartheta = \nabla\dot\vartheta = 0$ the phase gradient becomes stationary.
In a homogeneous system, without disorder $\sigma=0$, the constant phase profile $\vartheta(\mathbf{r})= \vartheta^0 =0$ solves the KPDE in synchronization, Eq.~(\ref{Eq:KPDEs02}), with $\Omega=\Omega^0=0$. In contrast, in the presence of heterogeneity $\sigma>0$ it is hard to obtain the  stationary phase profile because the synchronization frequency is not known and must be calculated self-consistently \cite{BlaToe05}.
Here we follow a perturbation approach by expanding in powers of the disorder $\sigma$.
Thereby, as will be shown below, non-trivial results are obtained in the second order.
\\ \\
Given the normalized frequencies $\eta(\mathbf{r})$ it is possible to derive the perturbation series
\beq	\label{Eq:OhmAnsatz01}
	\Omega(\sigma) ~=~ \sigma\Omega^{(1)} ~+~ \sigma^2\Omega^{(2)} ~+~ O(\sigma^3)
\eeq
directly by inserting the ansatz
\beq	\label{Eq:ThAnsatz01}
	\vartheta = \sigma\vartheta^{(1)} + \sigma^2\vartheta^{(2)} + O(\sigma^3)
\eeq
into Eq.~(\ref{Eq:KPDEs02}) and regrouping terms according to powers of $\sigma$
\beq	\label{Eq:OhmPert01}
	\Omega = \sum_{j=1}^\infty \sigma^j \left(\mathbf{L}\vartheta^{(j)} + \mathbf{b}^{(j)}\right) \,.
\eeq
Here, $\mathbf{L}=\nabla^2$ is the Hermitian, Laplacian operator, $\vartheta^{(j)}$ is the perturbation term of order $j$ in Eq. (\ref{Eq:ThAnsatz01}) and the functions $\mathbf{b}^{(j)}$ are given by
\beqarr	\label{Eq:Perterm00}
	b^{(1)}(\mathbf{r}) 	&=& \eta(\mathbf{r})	\nonumber \\ \nonumber \\
	b^{(2)}(\mathbf{r})	&=& \gamma\left[\nabla\vartheta^{(1)}(\mathbf{r})\right]^2 \\ \nonumber \\
	b^{(j>1)}(\mathbf{r})	&=& \gamma\sum_{i=1}^{j-1}\nabla\vartheta^{(i)}\nabla\vartheta^{(j-i)} 
	\quad. \nonumber
\eeqarr
To proceed, it is convenient to expand into eigenfunctions of the Laplacian $\mathbf{L}$.
For a medium of finite volume and appropriate boundary conditions the eigenvalues $E_k$ of $\mathbf{L}$ are discrete. With the orthonormal eigenfunctions $\mathbf{p}_k$ of the Laplacian and using the inner product of complex functions $\mathbf{f}$ and $\mathbf{g}$ defined for all positions $\mathbf{r}\in M$ of the medium
\beq
	\left(\mathbf{f}^\dagger\cdot\mathbf{g}\right) = \int_M d\mathbf{r} f^*(\mathbf{r}) g(\mathbf{r})
\eeq
we can define the projectors
\beq
	\mathbb{P}_0 = \mathbf{p}_0 \mathbf{p}_0^\dagger \qquad\textnormal{and}\qquad
	\mathbb{Q}_0 = \mathbb{I} - \mathbb{P}_0
\eeq
with the identity operator $\mathbb{I}$ and the constant function $p_0(\mathbf{r})=1/\sqrt{|M|}$ which is the normalized eigenfunction of $\mathbf{L}$ to the eigenvalue $E_0=0$. The operator $\mathbb{Q}_0$ removes, in fact, the average from a function. Applying these operators to Eq.~(\ref{Eq:OhmPert01}) we obtain
\beqarr
	\Omega^{(j)} 	&=& \mathbb{P}_0 \mathbf{b}^{(j)}	\label{Eq:Perterm01}	 \\ \nonumber \\
	0 		&=& \mathbf{L}\vartheta^{(j)} + \mathbb{Q}_0 \mathbf{b}^{(j)} \label{Eq:PertStep01}\quad.
\eeqarr
The inverse operator of $\mathbf{L}$ in the image space of $\mathbb{Q}_0$ is
\beq
	\mathbb{L}^{-1} = \sum_{k\ne 0} \frac{1}{E_k}\mathbf{p}_k\mathbf{p}_k^\dagger	\quad.
\eeq
We can thus solve Eq.~(\ref{Eq:PertStep01}) and find the perturbation terms $\vartheta^{(j)}$ up to a constant phase shift as
\beq	\label{Eq:Perterm02}
	\vartheta^{(j)} = 
	-\sum_{k\ne 0} {\frac{\left(\mathbf{p}_k^\dagger\cdot\mathbf{b}^{(j)}\right)}{E_k} \mathbf{p}_k}
	\quad.
\eeq
The equations (\ref{Eq:Perterm00}, \ref{Eq:Perterm01}, \ref{Eq:Perterm02}) can be iterated to obtain the 
full perturbation series (\ref{Eq:OhmAnsatz01}, \ref{Eq:ThAnsatz01}) up to arbitrary order.
Using the identities
\beqarr
	\left(\mathbf{p}_k^\dagger\nabla^\dagger\cdot\nabla \mathbf{p}_{k'}\right) &=& - E_k~\delta_{kk'}
	\\ \nonumber \\
	\left(\mathbf{p}_k^\dagger\cdot\eta\right)
	                \Bigl(\eta^\dagger\cdot\mathbf{p}_{k'}\Bigr)
	&=& \eta_k\eta_{k'} \label{Eq:CkkDef}
\eeqarr
and after some algebra we find for the first order and the second order perturbation term of the synchronization frequency
\beqarr
	\Omega^{(1)} &=& \bigl\langle \eta \bigr\rangle_\textnormal{System} \label{Eq:PertResult01} \\ \nonumber \\
	\Omega^{(2)} &=& - \gamma\frac{1}{|M|} \sum_{k\ne 0} \frac{\eta_k^2}{E_k}	\quad. \label{Eq:PertResult02}
\eeqarr
The coefficients $\eta_k^2$ are the square amplitudes of the $k^\textnormal{th}$ spatial Fourier modes of the frequencies, with respect to the system Laplacian.
For $k\ne 0$ these values do not depend on the mean value of $\eta(\mathbf{r})$. 
Note that for isochronous oscillations $\gamma=0$ the terms $\vartheta^{(j>1)}=0$ are zero and the phase profile in synchronization is given exactly by $\vartheta=\sigma\vartheta^{(1)}$ and Eqs.~(\ref{Eq:Perterm00}, \ref{Eq:Perterm02}). In that case, the phase diffusion equation (\ref{Eq:KPDEs01}) is linear and readily solved in the Fourier-Space.

\section*{III. Perturbation Approach 2}
In this section we will re-derive Eqs.~(\ref{Eq:PertResult01},\ref{Eq:PertResult02}) from a different point of view and in a somewhat more general form. 
It is well known that a nonlinear Cole-Hopf transformation
\beq	\label{Eq:ColeHopf01}
	\vartheta(\mathbf{r}) = \frac{1}{\gamma}\ln p(\mathbf{r})
\eeq
changes the KPDE~(\ref{Eq:KPDEs02}) into a linear equation
\beq	\label{Eq:Schroed01}
	\Omega\gamma~p(\mathbf{r}) 	~=~ \left[\gamma\sigma~\eta(\mathbf{r}) ~+~ \nabla^2\right]~p(\mathbf{r})
					~=~ -\mathbf{H}~p(\mathbf{r})
\eeq
for the ground state $p_0(\mathbf{r})=p(\mathbf{r})$ of a Hamiltonian $\mathbf{H}$ with diagonal disorder, given by the frequencies $-\gamma\sigma\eta(\mathbf{r})$, and ground state energy $-\gamma\Omega$ (see,~e.g.~\cite{Kuramoto84,Sakaguchi88}). Considering the frequencies $\eta(\mathbf{r})$ a perturbation of strength $\varepsilon=\gamma\sigma$, Schr\"odinger perturbation theory will give exactly the same results for the synchronization frequency as obtained in the previous section Eqs.~(\ref{Eq:PertResult01},\ref{Eq:PertResult02}). 
However, as will be shown below the stationary Schr\"odinger equation is only one from a family of linear problems which are equivalent to the KPDE.
\\ \\
To understand this, notice that there are two pitfalls to the transformation Eq.~(\ref{Eq:ColeHopf01}). First, it is only defined for non vanishing values $\gamma\ne 0$. 
And secondly, since the ground state $p_0(\mathbf{r})=1/\sqrt{|M|}$ of the unperturbed system Eq.~(\ref{Eq:Schroed01}) is unique one is tempted to believe that the same is true for the homogeneous phase profile ($\vartheta=\textnormal{const}$) of identical oscillators in synchronization. However, this is not necessarily the case because the phases $\vartheta(\mathbf{r})$ are elements of a circle while $p(\mathbf{r})$ is a real number. If the phase changes along a closed path 
from zero to a multiple of $2\pi$ it is a continuous function on this curve 
while $\mathbf{p}$ is necessarily discontinuous. Indeed, for periodic boundary conditions multiple stable synchronized states can exist \cite{Strogatz06} (we will give examples for such inhomogeneous solutions in the next Section).
\\ \\
We will therefore not take $\vartheta^0(\mathbf{r})=0$ as in the previous section,
but instead assume a general phase profile $\vartheta^0(\mathbf{r})$ in stable synchronization.
Note that it is always possible to divide the phases formally into a time independent `gauge' field $\vartheta^0$ and a time dependent deviation $\varphi$ from that gauge field, $\vartheta(\mathbf{r},t)=\vartheta^0(\mathbf{r})+\varphi(\mathbf{r},t)$ . 
This corresponds to local rotations (i.e., the gauge field $\vartheta^0(\mathbf{r})$ defines a position dependent choice of the coordinate frame) and yields new phases $\varphi(\mathbf{r})$, so that $\varphi=0$ where in the old frame $\vartheta=\vartheta^0$. 
%
We can define 
\beq
\label{Eq:HomSol01}
	\Omega^0(\mathbf{r}) = \nabla^2\vartheta^0 + \gamma\left(\nabla\vartheta^0\right)^2 \quad.
\eeq
Then the KPDE of the full heterogeneous system 
in synchrony take the new form
\beq
	\Omega = \omega(\mathbf{r}) + \Omega^0(\mathbf{r}) + \nabla^2\varphi + 2\gamma\nabla\vartheta \nabla\varphi + \gamma\left(\nabla\varphi\right)^2
\eeq
and the gauge modified Laplacian reads
\beq
\label{Eq:GaugeModLapl}
\mathbf{L} = 2\gamma \nabla\vartheta^0\nabla + \nabla^2 \,.
\eeq
After the Cole-Hopf transformation $\varphi(\mathbf{r}) =\frac{1}{\gamma}\log p(\mathbf{r})$ we find the eigenvalue problem
\beq	\label{Eq:SchroedGauge01}
	\gamma\Omega~\mathbf{p} ~=~ \left[\gamma\sigma\eta(\mathbf{r}) + \Omega^0(\mathbf{r})
								+ 2\gamma\nabla\vartheta^0\nabla
								+ \nabla^2 \right] \, \mathbf{p} \quad.
\eeq
The change of gauge is nothing else but a similarity transformation of the Hamiltonian and does not effect its eigenvalues, unless the gauge field itself contains topological charges. 
If we choose a synchronized solution of Eq.~(\ref{Eq:KPDEs02}) as gauge field so that $\Omega^0(\mathbf{r})=\Omega^0$ is constant Eq.~(\ref{Eq:SchroedGauge01}) can be written as
\beq	\label{Eq:SchroedGauge02}
	\gamma\left(\Omega-\Omega^0\right)~\mathbf{p} ~=~ \left[\gamma\sigma\mathbf{V}_\eta
								+ 2\gamma\nabla\vartheta^0\nabla
								+ \nabla^2 \right]~ \mathbf{p} \quad.
\eeq
where a non-Hermitean operator $\mathbf{L}=2\gamma\nabla\vartheta^0\nabla + \nabla^2$~ is perturbed by a diagonal disorder $\varepsilon\mathbf{V}_\eta=-\varepsilon~\textnormal{diag}(\eta)$ with strength $\varepsilon=\gamma\sigma$. Note that the constant function $p_0(\mathbf{r})=1/\sqrt{|M|}$ is an eigenfunction of $\mathbf{L}$ to the zero eigenvalue $E^0_0=0$. This corresponds to a constant phase shift from $\vartheta^0(\mathbf{r})$, i.e. the stable synchronization manifold for which we seek a perturbation expansion.
Given orthonormal left and right eigenfunctions $\mathbf{P}_k$ and $\mathbf{p}_k$ of $\mathbf{L}$ and the corresponding eigenvalues $E_k^0$, the terms in the perturbation series for the eigenvalue $E_0$ of $-\mathbf{\tilde H}$ with the largest real part
\beq
	\gamma\left(\Omega-\Omega^0\right) ~=~ E_0 ~=~ \varepsilon E_0^{(1)} ~+~ \varepsilon^2 E_0^{(2)} ~+~ O(\varepsilon^3) 
\eeq
are found to be
\beqarr
	E_0^{(1)}	&=& \left(\mathbf{P}_0^\dagger\cdot\mathbf{V}_\eta\mathbf{p}_0\right)	\nonumber \\ \label{Eq:PertResult003}\\
      E_0^{(2)}	&=& - \sum_{k\ne 0} \frac{	\left(\mathbf{P}_0^\dagger\cdot\mathbf{V}_\eta\mathbf{p}_k\right)
						\left(\mathbf{P}_k^\dagger\cdot\mathbf{V}_\eta\mathbf{p}_0\right) }{E_k^0}	
						\nonumber \qquad.
\eeqarr
The discrimination between left and right eigenfunctions is necessary because $\mathbf{L}$ is not Hermitian unless $\nabla\vartheta^0=0$. For certain regular topologies and symmetric solutions $\vartheta^0$ the left and right eigenfunctions $\mathbf{P}_k=\mathbf{p}_k$ are identical, nevertheless. In this case, we obtain the perturbation terms for the synchronization frequency as
\beqarr
	\Omega 	&=& 	\Omega^0 + \sigma \Omega^{(1)} + \sigma^2 \Omega^{(2)} + O(\sigma^3) \label{Eq:OhmAnsatz02}\\ \nonumber \\
	\Omega^{(1)} 	&=&	\bigl\langle\eta\bigr\rangle_\textnormal{System} \label{Eq:PertResult03} \\ \nonumber \\
      \Omega^{(2)} &=&	- \gamma \frac{1}{|M|}\sum_{k\ne 0} \frac{\eta_k^2}{E_k^0}	\label{Eq:PertResult04}
\eeqarr
with the spatial Fourier modes $\eta_k$ of the frequencies.
%
This result can directly be compared to Eqs.~(\ref{Eq:OhmAnsatz01},\ref{Eq:PertResult01},\ref{Eq:PertResult02}). The first noticeable difference is, that the unperturbed system may have a synchronization frequency $\Omega^0$ which is different from zero. The second difference is more subtle. The eigenvalues $E_k^0$ may have non vanishing imaginary parts. This corresponds to oscillatory modes during the transient to synchronization. Nevertheless, the sum Eq.~(\ref{Eq:PertResult04}) and perturbation expansion Eq.~(\ref{Eq:OhmAnsatz02}) are real if eigenvalues and eigenfunctions occur in complex conjugated pairs.

\section*{IV. Examples}
\subsection*{Solution in a rectangular medium}
\begin{figure}[tb]
\center
\includegraphics[width=6.75cm]{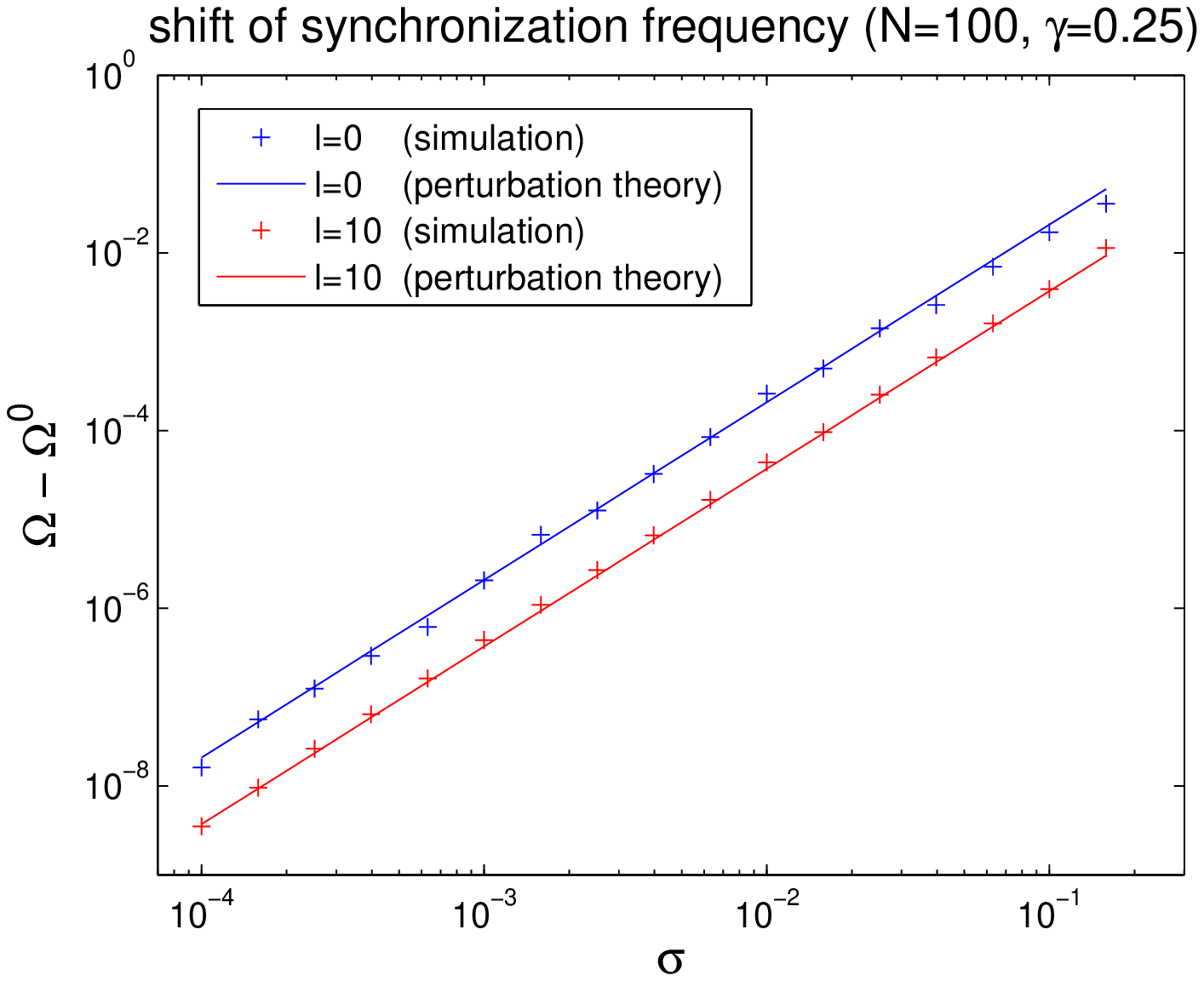}
\includegraphics[width=6.75cm]{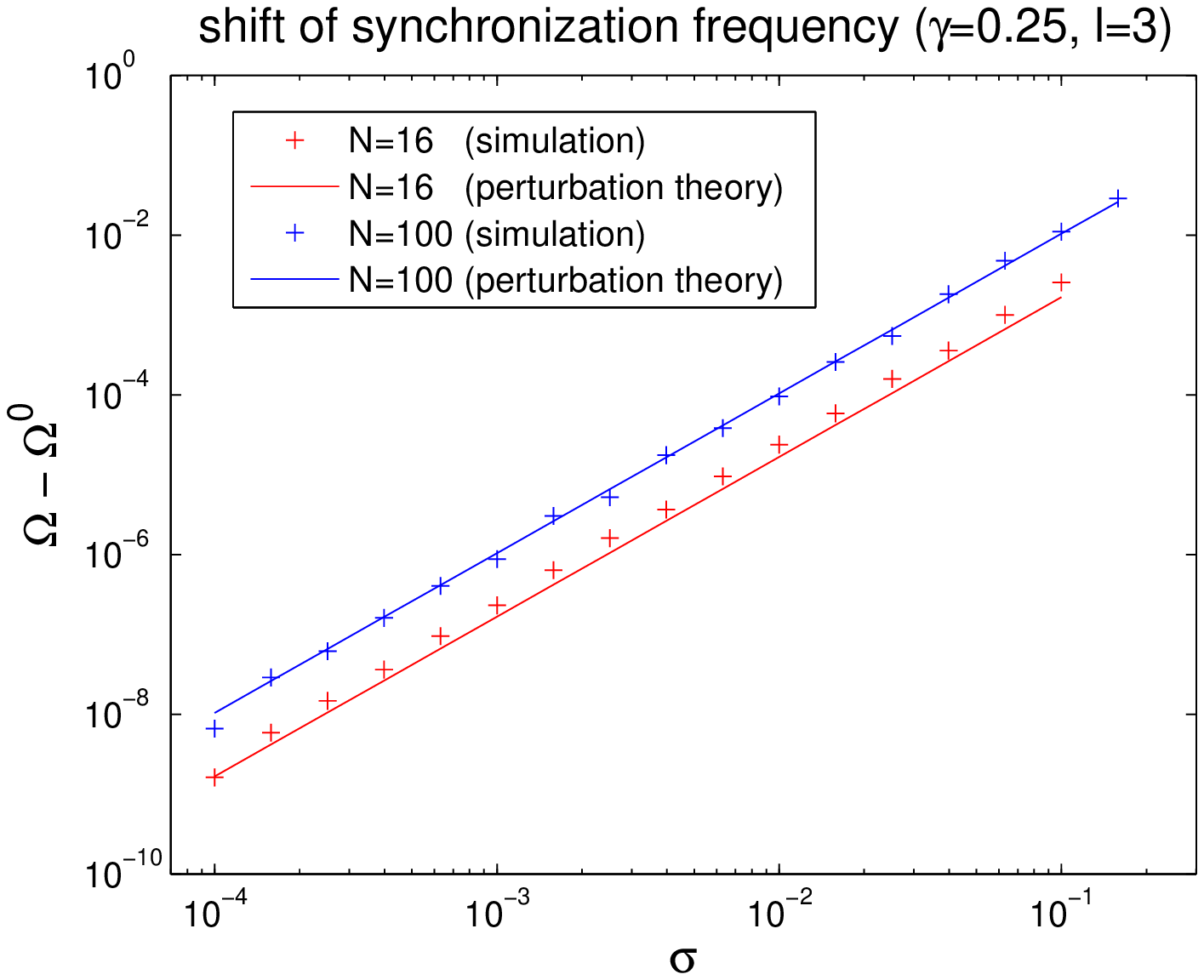}
\caption{\small 
Influence of frequency disorder in a one-dimensional lattice.
Plotted is the shift of the synchronization frequency $\Omega$ from the frequency of the synchronized state with identical oscillators, $\Omega_0$, as a function of the oscillator heterogeneity $\sigma$. We compare simulations  
of the discrete Kuramoto model Eq.~(\ref{Eq:KPEs01}) for a chain of $N$ oscillators (crosses) with the second order perturbation theory (solid lines), Eq.~(\ref{Eq:PertResult05}) with $L\to N$. 
Each cross from the simulations is an average value from $50$ runs with different realizations of iid.~random frequencies ($C_{kk}=1$), where
for each realization the average frequency has been shifted to zero $\bigl\langle\eta\bigr\rangle_\textnormal{System}=0$.
Left: comparison of the results for a ring of $N=100$ nonisochronous ($\gamma=0.25$) phase oscillators, without topological charge (blue graph) and with a topological charge of $l=10$ (red graph). 
Right: comparison of rings of different sizes $N=16$ (red graph) and $N=100$ (blue graph) oscillators but with the same nonisochronicity $\gamma=0.25$ and topological charge $l=3$.}
\label{FigFreqShift1d}
\end{figure}

In the following we are interested to apply these results to some simple topologies.
In order to apply our perturbation approach the spectrum of the Laplacian has to be calculated for every topology of interest. We start by examining a simple lattice.
Let us consider a $d$-dimensional oscillatory medium $L^d \subset \mathbb{R}^d$ with periodic boundary conditions and quenched random frequency disorder. We first note, that a constant phase gradient $\nabla\vartheta^0 = \frac{2\pi}{L}\textbf{\em l}$~, where $\textbf{\em l}=(l_1,\dots,l_d)$~ is a $d$-dimensional integer vector of winding numbers, solves the homogeneous equation Eq.~(\ref{Eq:HomSol01}) with
\beq
	\Omega^0 = \gamma\left(\frac{2\pi}{L}\right)^2 |\textbf{\em l}\,|^2	\quad.
\eeq
Because of the periodic boundary conditions of the medium and the phases $\vartheta=\vartheta+2\pi$ we can include topological charges without phase singularities. The phase gradient is bounded and the use of the Kuramoto phase diffusion equation is justified. 
The unperturbed operator, Eq.~(\ref{Eq:GaugeModLapl}), reads
\beq
	\mathbf{L} = 2\gamma\frac{2\pi}{L}\textbf{\em l}^{\dagger}\nabla + \nabla^2
\eeq
and the left and right eigenfunctions and eigenvalues that fulfill the periodic boundary conditions coincide and are simple harmonics
\beqarr	
	p_\mathbf{k}(\mathbf{r}) &=& L^{-\frac{d}{2}} e^{i\frac{2\pi}{L}\mathbf{k}^\dagger\cdot\mathbf{r}} 
	\label{Eq:Eigf01} \\ \nonumber \\
\mathbf{L}~\mathbf{p}_\mathbf{k} &=& \left(\frac{2\pi}{L}\right)^2 (i~2\gamma \textbf{\em l}^{\dagger}\cdot\mathbf{k} - |\mathbf{k}|^2)
							~\mathbf{p}_\mathbf{k}	~=~ E_\mathbf{k}^0~\mathbf{p}_\mathbf{k}\qquad.
	\label{Eq:Eigv01}
\eeqarr
The vector $\mathbf{k}=(k_1\dots,k_d)$ is also an integer vector, labeling the Fourier modes in the various directions. 
Eq.~(\ref{Eq:PertResult04}) for the second order correction of the synchronization frequency shift gives
\beq	\label{Eq:PertResult05}
	\Omega^{(2)} = \gamma L^{2-d} \frac{1}{4\pi^2} \sum_{|\mathbf{k}|\ne 0} 
	\frac{|\mathbf{k}|^2}{4\gamma^2(\textbf{\em l}^\dagger\cdot\mathbf{k})^2+|\mathbf{k}|^4}~ \eta_\mathbf{k}^2	\quad.
\eeq
This sum over the $d$-dimensional integer lattice is potentially divergent depending on the small wavelength behavior of the terms $\eta_\mathbf{k}^2$. Delta correlated random frequencies lead to an ultraviolet divergence in dimensions $d>1$ larger than one. We therefore have to restrict the perturbation theory to cases, where nearby frequencies are correlated, for instance as  
\beq	\label{Eq:CorrAnsatz01}
	\mathbb{E}\left[\eta(\mathbf{r})\eta(\mathbf{r}')\right] - \mathbb{E}\left[\eta\right]^2
	~=~ \left(2\pi\lambda^2\right)^{-\frac{d}{2}}~e^{-\frac{|\mathbf{r}-\mathbf{r}'|^2}{2\lambda^2}} 
	\eeq
and $\lambda$ is some correlation length. Note, that Eq.~(\ref{Eq:CorrAnsatz01}) can only be an approximation for small correlation lengths compared with the system size $L$, disregarding boundary effects.  The expected value of the Fourier coefficients for $|\mathbf{k}|\ne 0$ is then
\beq	\label{Eq:CorrAnsatz02}
	\mathbb{E}\left[\eta_\mathbf{k}^2 \right] = e^{-2\left(\frac{\lambda\pi}{L}\right)^2|\mathbf{k}|^2}	\quad.
\eeq
Using this expression one can calculate the expected second order perturbation terms in Eq.~(\ref{Eq:PertResult05}) numerically. An exact analytical expression exists in the simplest case of $d=1$, $l=0$ and delta correlated frequencies with $\lambda=0$. Then expression (\ref{Eq:PertResult05}) becomes
\beq
	\mathbb{E}\left[\Omega^{(2)}\right] = \gamma L \frac{1}{2\pi^2} \sum_{k>0} 
	\frac{1}{k^2} = \gamma \frac{L}{12}
\eeq
where we could use the property of the Riemann zeta-function $\zeta(2)=\pi^2/6$. 
In Fig.~\ref{FigFreqShift1d} we compare the shift of the synchronization frequency due to heterogeneity for one dimensional systems with periodic boundary conditions, different lengths and winding numbers $l$.
In order to observe the second order terms the linear contribution to the frequency shift must be exactly zero. This is achieved by shifting the average frequencies to zero $\bigl\langle\eta\bigr\rangle_\textnormal{System}=0$ for each realization. The second order perturbation term is not affected by this change into a co-rotating frame of reference. 
The figure confirms the asymptotic behavior of the synchronization frequency for $\gamma\sigma\ll 1$ and
we find the scaling relation 
\beq
\Omega - \Omega_0 \sim \gamma\sigma ^ 2
\eeq
as was previously observed in \cite{BlaToe05}.
\\ \\
The scaling of the second order perturbation term with the system size $L$ and the correlation length $\lambda$ can be studied by approximating the sum with a $d$-dimensional integral over $|\mathbf{k}|\ge 1$ 
%
%
\beqarr	\label{Eq:Scaling01}
	\mathbb{E}\left[\Omega^{(2)}\right] &\sim& \gamma L^{2-d} \int_{|\mathbf{k}|>1} |\mathbf{k}|^{-2} 
		e^{-2\left(\frac{\lambda\pi}{L}\right)^2|\mathbf{k}|^2} d\mathbf{k} \nonumber \\ \\
	&\sim& \gamma~L^{2-d}~x^{2-d}~
	\Gamma\left[\frac{d-2}{2}~,~x^2\right] \quad\textnormal{with}\quad x=\frac{\lambda}{L}\pi\sqrt 2\quad. \nonumber
\eeqarr
We have here omitted constant factors, for instance from the integration over the $d$-dimensional sphere shells $|\mathbf{k}|=\mathrm{const}$ or upper and lower bounds that allow the estimation of the sum from an integral. Depending on the dimension $d$ we can use different asymptotic scaling relations of the incomplete gamma-function in Eq.~(\ref{Eq:Scaling01}) for $x\to 0$, i.e. large system sizes or small correlation lengths. We find
\beqarr	\label{Eq:Scaling02}
	\mathbb{E}\left[\Omega^{(2)}\right] &\sim & O\left( L \right)	\qquad\qquad\qquad\textnormal{for}\quad d=1
	\nonumber \\ \nonumber \\
	\mathbb{E}\left[\Omega^{(2)}\right] &\sim& O\left( \log\left(\frac{L}{\lambda}\right)\right)	\qquad\textnormal{for}\quad d=2	
	\\ \nonumber \\
	\mathbb{E}\left[\Omega^{(2)}\right] &\sim& O\left( \lambda^{2-d}\right)	\qquad\qquad~\textnormal{for}\quad d\ge 3	\quad. 
	\nonumber
\eeqarr
The analysis for a rectangular medium with no-flux or open boundary conditions gives analogous results, with the only difference that no topological charges are possible, i.e.~$|\textbf{\em l}|=0$.

\subsection*{Solution on a sphere}

\begin{figure}[tb]
\center
	\mbox{
	(a)~\includegraphics[width=4cm]{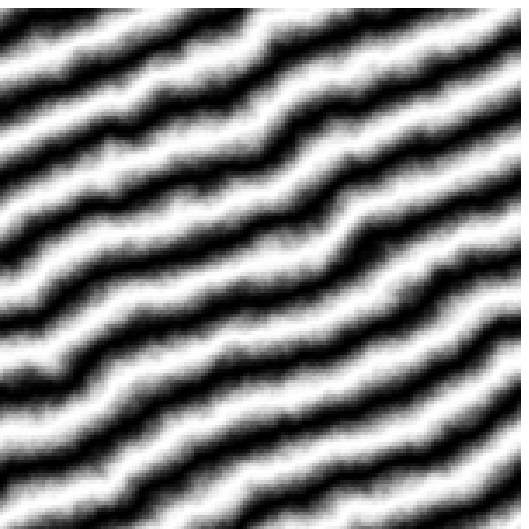}
	\qquad(b)~\includegraphics[width=6.3cm]{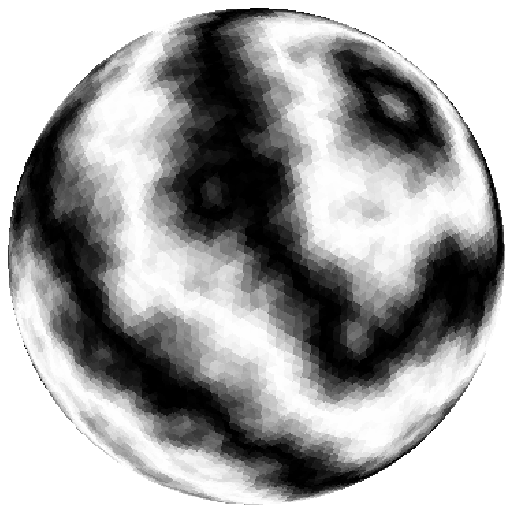}
	}
\caption{\small Quasi-regular wave patterns (a) in a rectangular medium with periodic boundary conditions and (b) on the surface of a sphere. We have used the discrete Kuramoto model Eq.~(\ref{Eq:KPEs01}) on (a) a grid of $150\times 150$ phase oscillators and (b) on an almost homogeneous discretization of the sphere surface with $20480$ points on the faces of a triangular tessellation \cite{BaumFred85}. The natural frequencies of the individual oscillators were independently uniformly distributed with standard deviation $\sigma=0.2$. The topology of the square lattice in subfigure (a) is that of a 2-torus and it is possible to have topological charges without phase singularities (large phase differences). We used an initial condition with topological charges $l_x=3$ and $l_y=7$. Shown is the sine of the phases in gray levels after a transient time to synchronization.}
\label{FigSurfaceWaves}
\end{figure}

Of special interest may be the synchronization frequency of a heterogeneous, oscillatory, reaction diffusion system on the surface of a sphere as a model for catalytic surface reactions on spherical bodies. Unlike in the torus topology of a rectangular medium with periodic boundary conditions, on the sphere topological charges always occur in vortex pairs of opposite charge. The method of phase reduction is not applicable in the vicinity of such phase singularities which act as fast pacemakers for the system. We will therefore only study perturbations of the homogeneous synchronized solution $\vartheta^0=0$. The eigenfunctions of the Laplacian on a sphere of radius $R$ are spherical harmonics $\mathbf{Y}_{lm}$ with
\beqarr
	\mathbf{p}_{lm} &=& \frac{1}{R}\mathbf{Y}_{lm}
	\qquad\textnormal{with}\quad l=0,1,\dots \quad\textnormal{and}\quad m=-l,\dots, l	\nonumber \\ \\
	\nabla^2 \mathbf{p}_{lm} &=& -\frac{1}{R^2} l(l+1) \mathbf{p}_{lm} 	\qquad.	\nonumber 
\eeqarr
If we assume a homogeneous, isotropic distribution of frequencies on the sphere, the frequency correlator must have an $SO(3)$ symmetry. Let $\mathbf{U}_{\mathbf{r}}$ be a transformation with $\mathbf{U}_{\mathbf{r}}\mathbf{e}_{\mathbf{z}} = \mathbf{e}_{\mathbf{r}}$, which first rotates a vector in $z$-direction $\mathbf{e}_{\mathbf{z}}$ around the $y$-axis to the zenith of $\mathbf{e}_{\mathbf{r}}$ and subsequently around the $z$-axis to its azimuth. If we make the ansatz for an isotropic, homogeneous correlation kernel
\beq
	 \mathbb{E}\left[\eta(\mathbf{r})\eta(\mathbf{r}')\right] - \mathbb{E}\left[\eta\right]^2 
         = \frac{1}{R^2}\sum_{l=0}^\infty c_l~ \sqrt{\frac{2l+1}{4\pi}}~ Y_{l0}\left(\mathbf{U}_{\mathbf{r}}^{-1}\mathbf{r}'\right)
\eeq
we find the expected values of the spherical harmonics spectrum of the quenched frequency disorder as
\beq
	\mathbb{E}\left[\eta_{lm}^2\right] = c_l \,.
\eeq
If the frequencies $\eta(\mathbf{r})$ are delta correlated all coefficients $c_{l}$ with $l>0$ are equal to one and the sums 
\beq
	\sum_{l=1}^\infty\sum_{m=-l}^l \frac{C_{lm}}{l(l+1)} = \sum_{l=1}^\infty c_l~\frac{2l+1}{l(l+1)}
\eeq
in Eqs.~(\ref{Eq:PertResult02},\ref{Eq:PertResult04}) are divergent. We, therefore, have to assume a cutoff at a wave number $l_{\max}\sim R/\lambda$ where $\lambda$ is a correlation length. This cutoff can be sharp or exponential, as in the previous example, and we obtain again a relation
\beq
	\mathbb{E}\left[\Omega^{(2)}\right] \sim O\left(\log\left(\frac{R}{\lambda}\right)\right) \,.
\eeq

\section*{V. Failure of Perturbation Theory}
\begin{figure}[!htp]
	\center
	\mbox{
	\includegraphics[width=6.4cm]{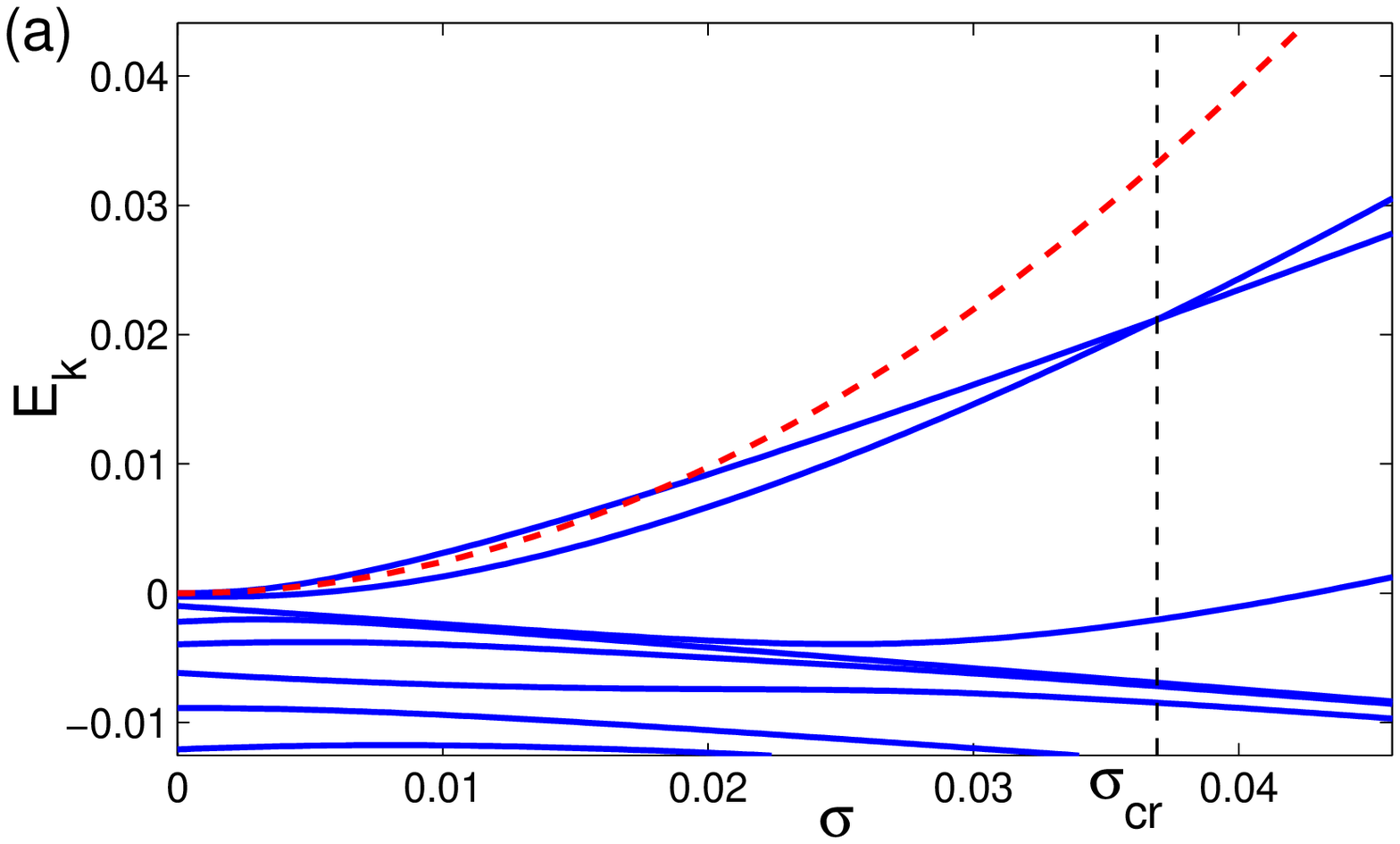} 
	\includegraphics[width=6.4cm]{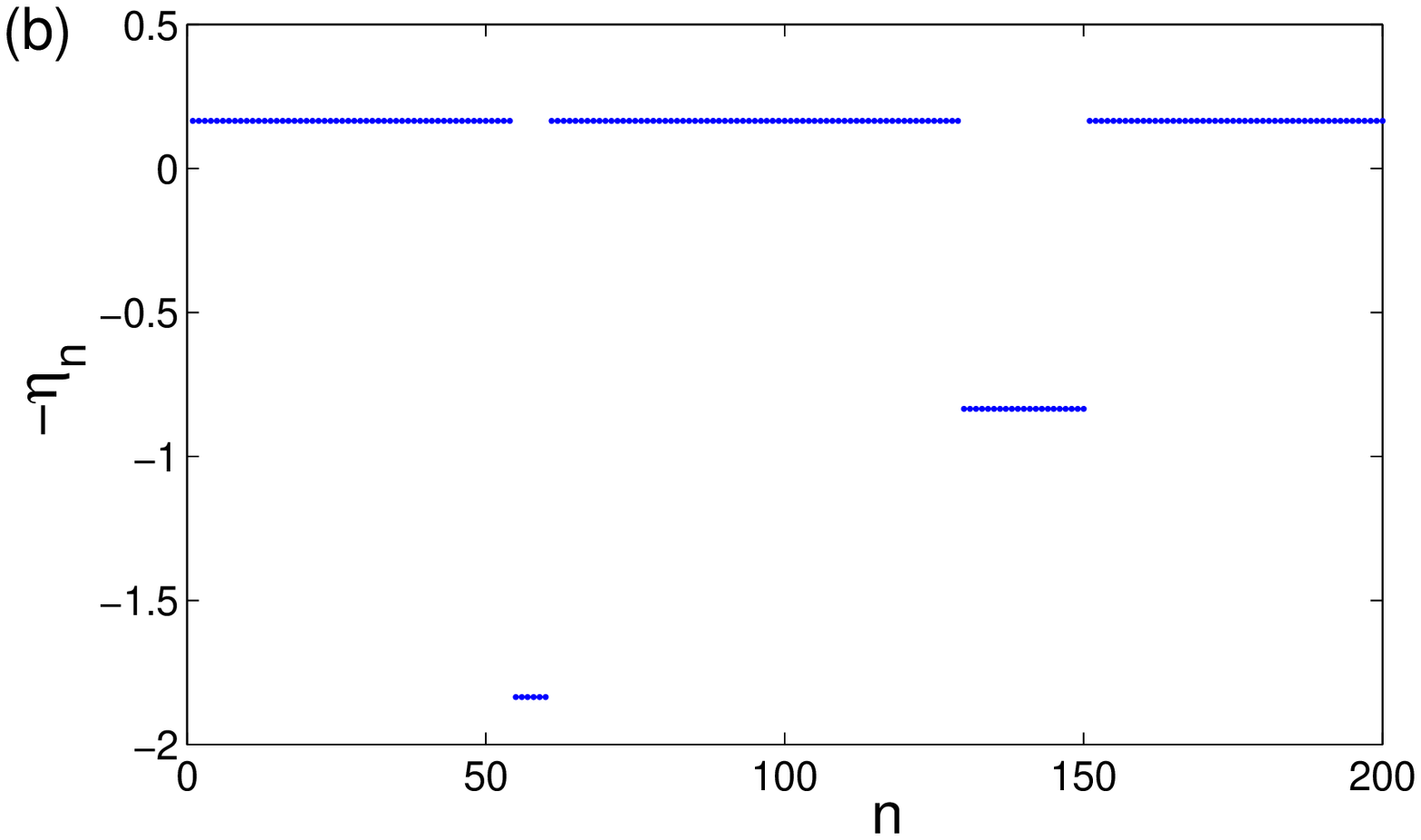}
	} \\
	\mbox{
	\includegraphics[width=6.4cm]{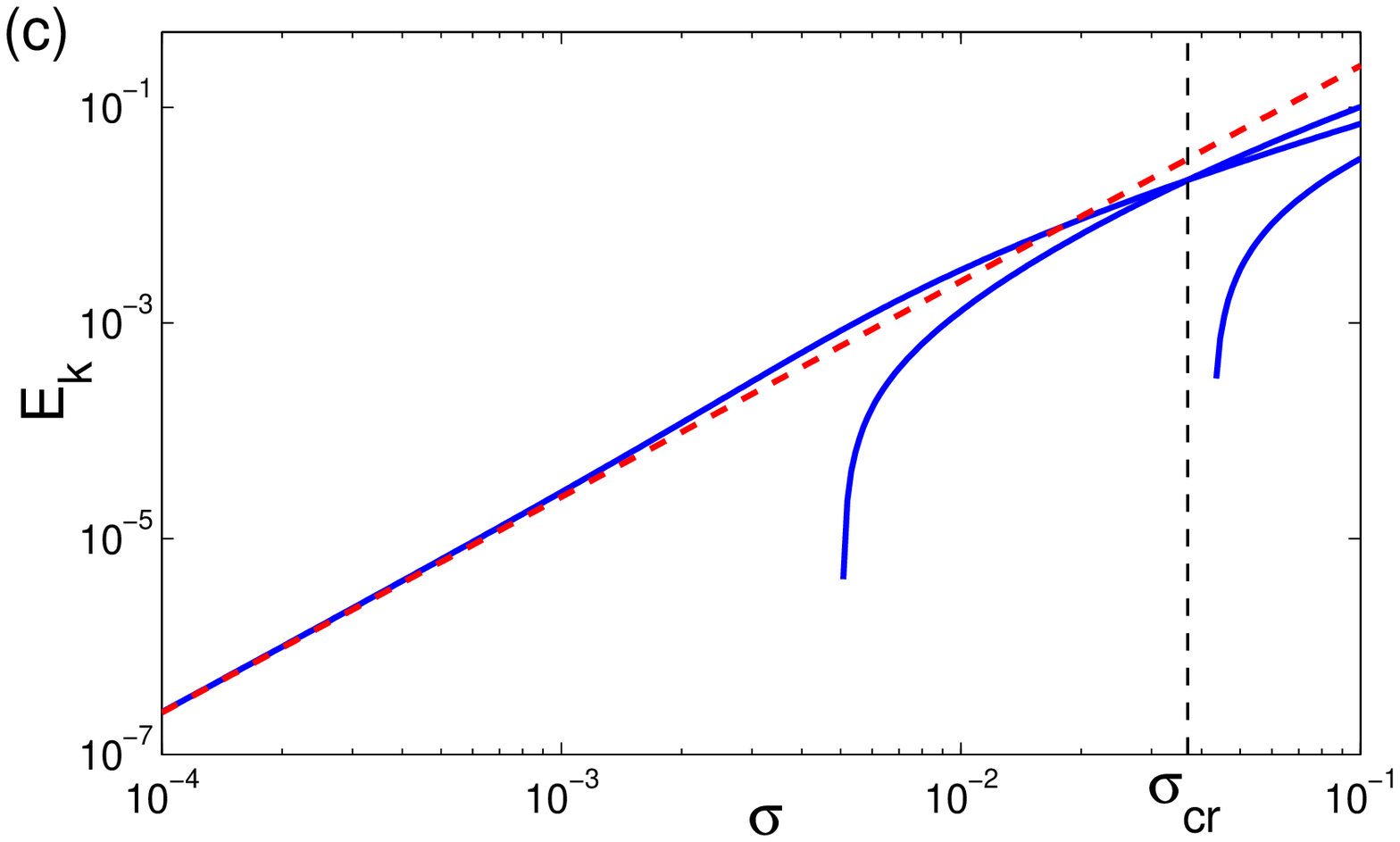}
	\includegraphics[width=6.4cm]{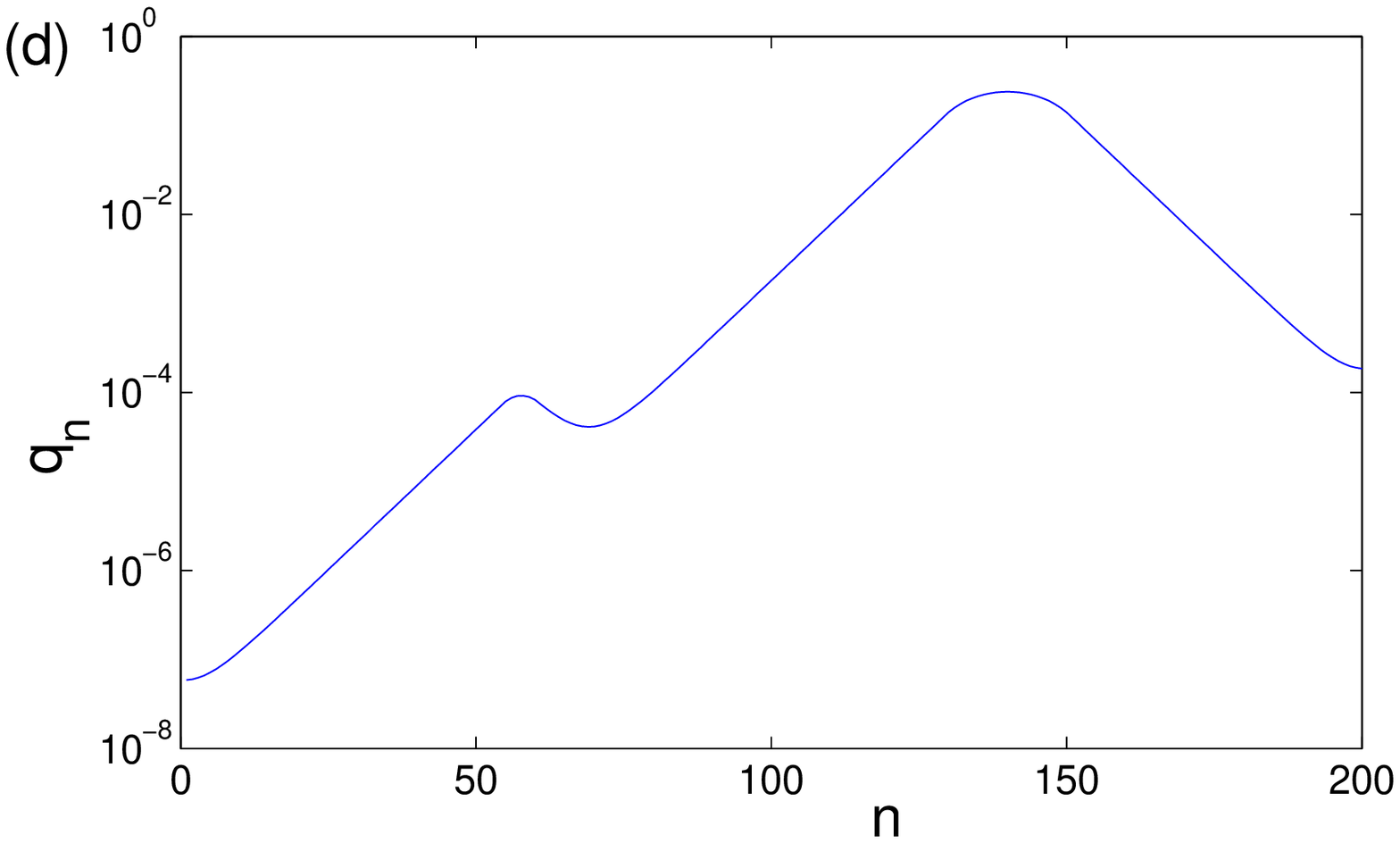}
	} \\
	\mbox{
	\includegraphics[width=6.4cm]{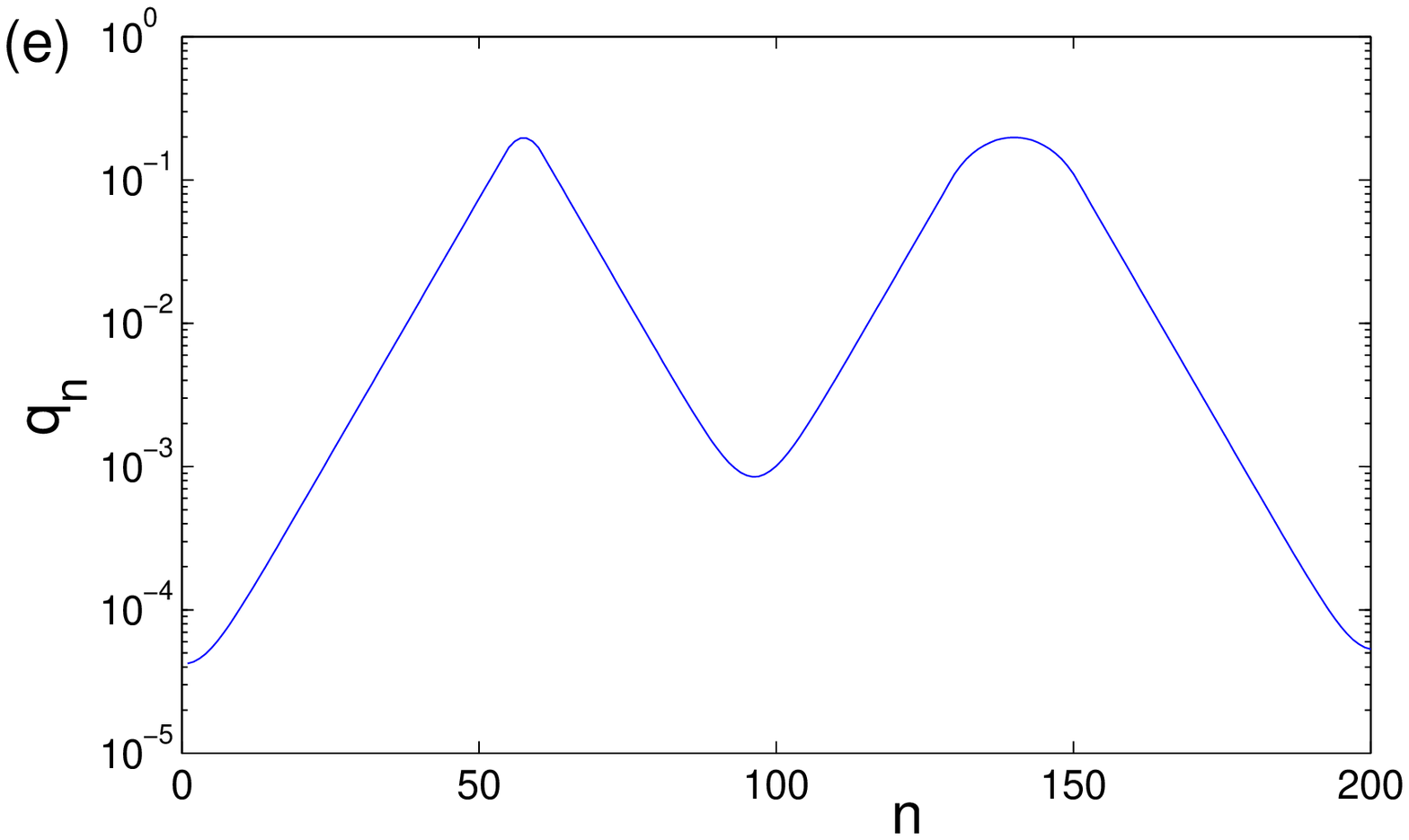}	
	\includegraphics[width=6.4cm]{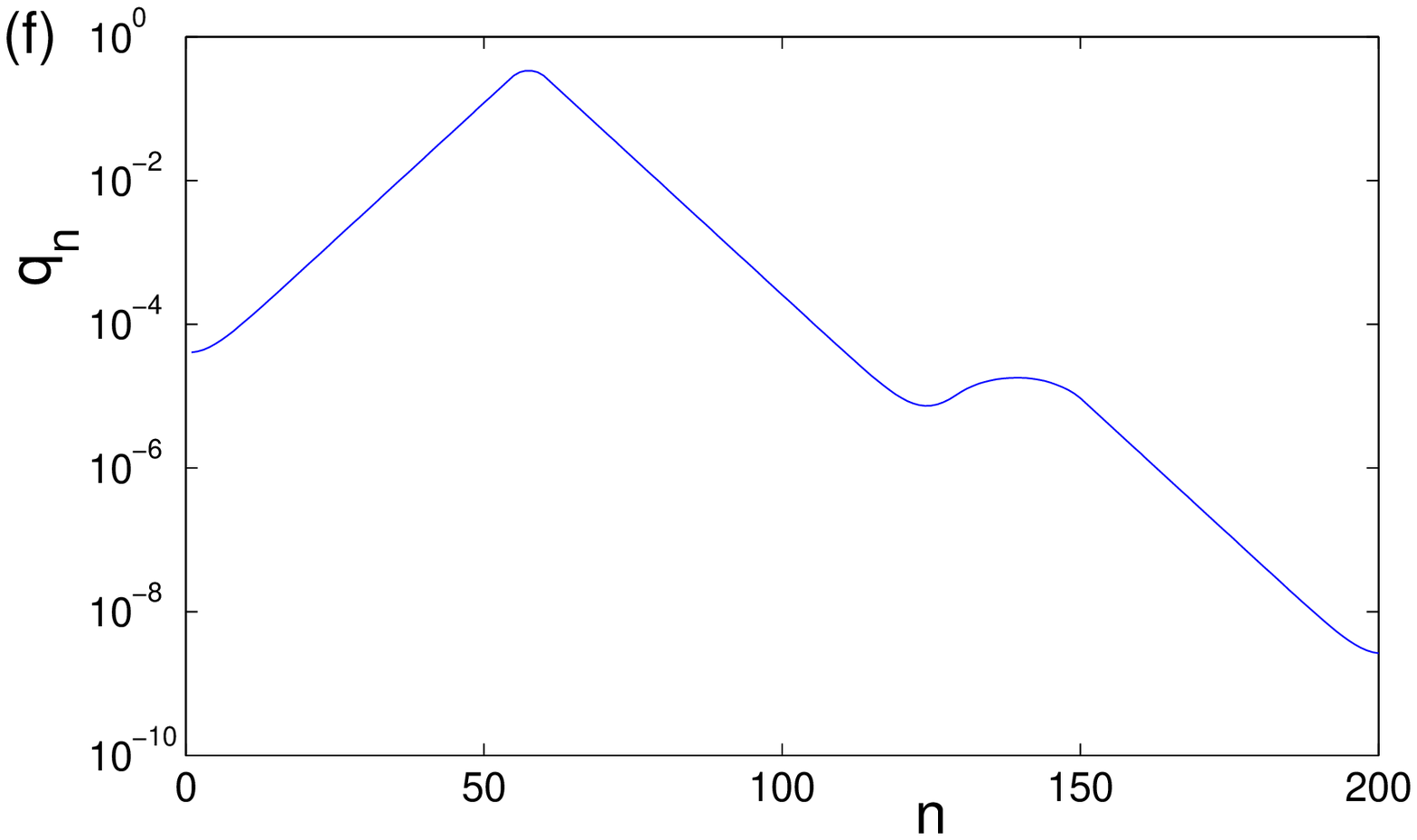}
	}              
\caption{\small 
Synchronization analysis in a one dimensional chain of $N=200$ nonidentical Kuramoto phase oscillators with two competing pacemakers regions according to Eq.~(\ref{Eq:diskr1}).
(b) frequency values $-\eta_n$ corresponding to the potential of the discrete Hamiltonian  (values $\eta_n$ are shifted so that the mean $\bigl\langle\eta\bigr\rangle_\textnormal{System}=0$ is exactly zero). There are two potential wells at which the ground state can be localized, a deeper well on the left and a shallower but broader well on the right. (a) and (c) largest eigenvalues (solid blue lines) of the negative Hamiltonian $-\mathbf{H}$ in dependence on the heterogeneity $\sigma$, the second order perturbation approximation (Eq.~(\ref{Eq:PertResult003}), dashed red line), and the value $\sigma_{cr}$ for which the ground state becomes quasi degenerate (dashed black line). The quality of the approximation can be seen in double logarithmic scales in subfigure (c). (d)-(f) numerically determined groundstate eigenvectors in a semilogarithmic scale for (d) $\sigma=0.03$, (e) $\sigma=\sigma_{cr}=0.036917$ near the point of quasi degeneracy and (f) $\sigma=0.04$. Exponential localization at a potential well corresponds to concentric waves around this pacemaker region in the Kuramoto phase diffusion equations.}
	\label{fig:DoubleCrossing}
\end{figure}

\begin{figure}[!htp]
	\center
	\mbox{
	\includegraphics[width=6.4cm]{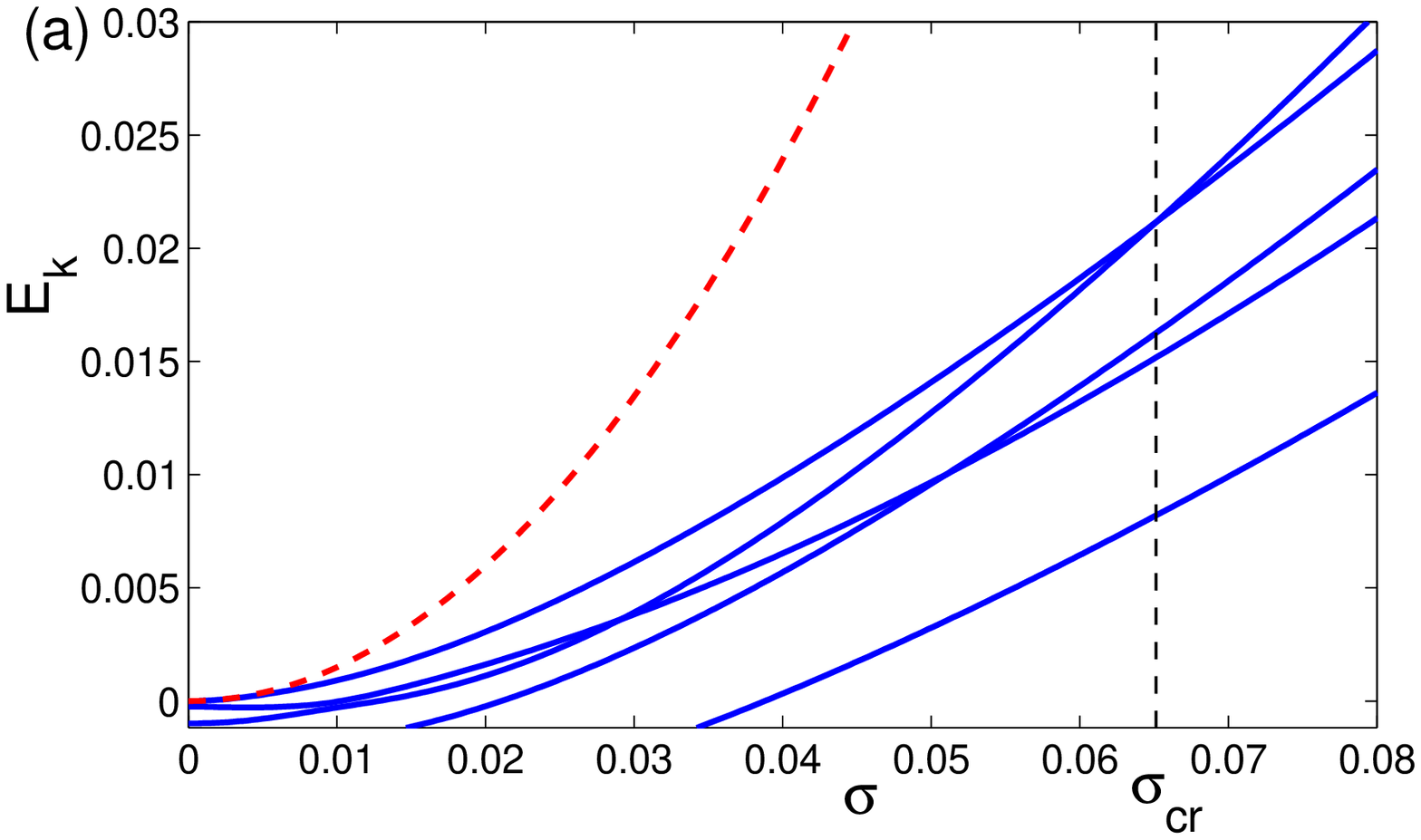} 
	\includegraphics[width=6.4cm]{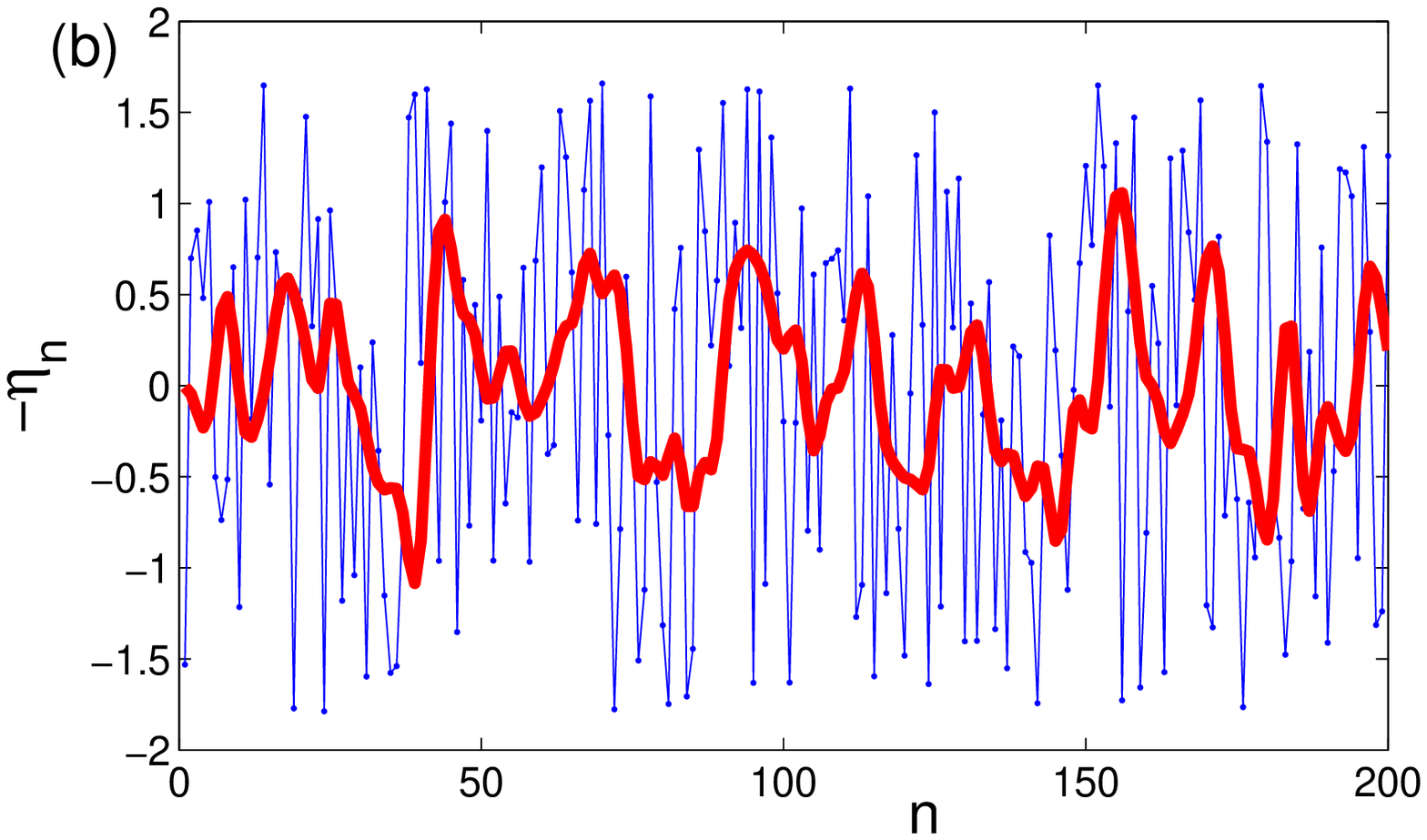}
	} \\
	\mbox{
	\includegraphics[width=6.4cm]{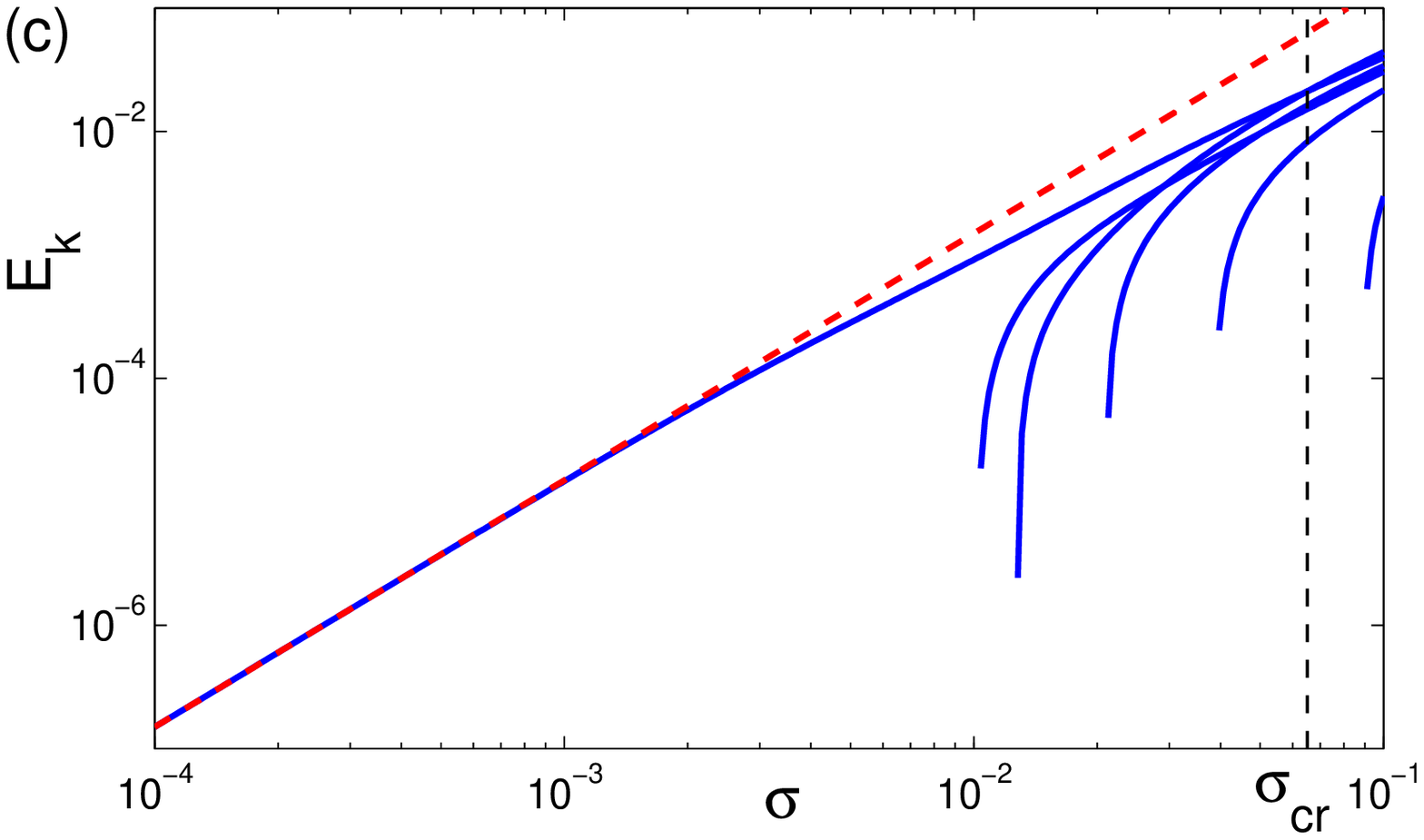}
	\includegraphics[width=6.4cm]{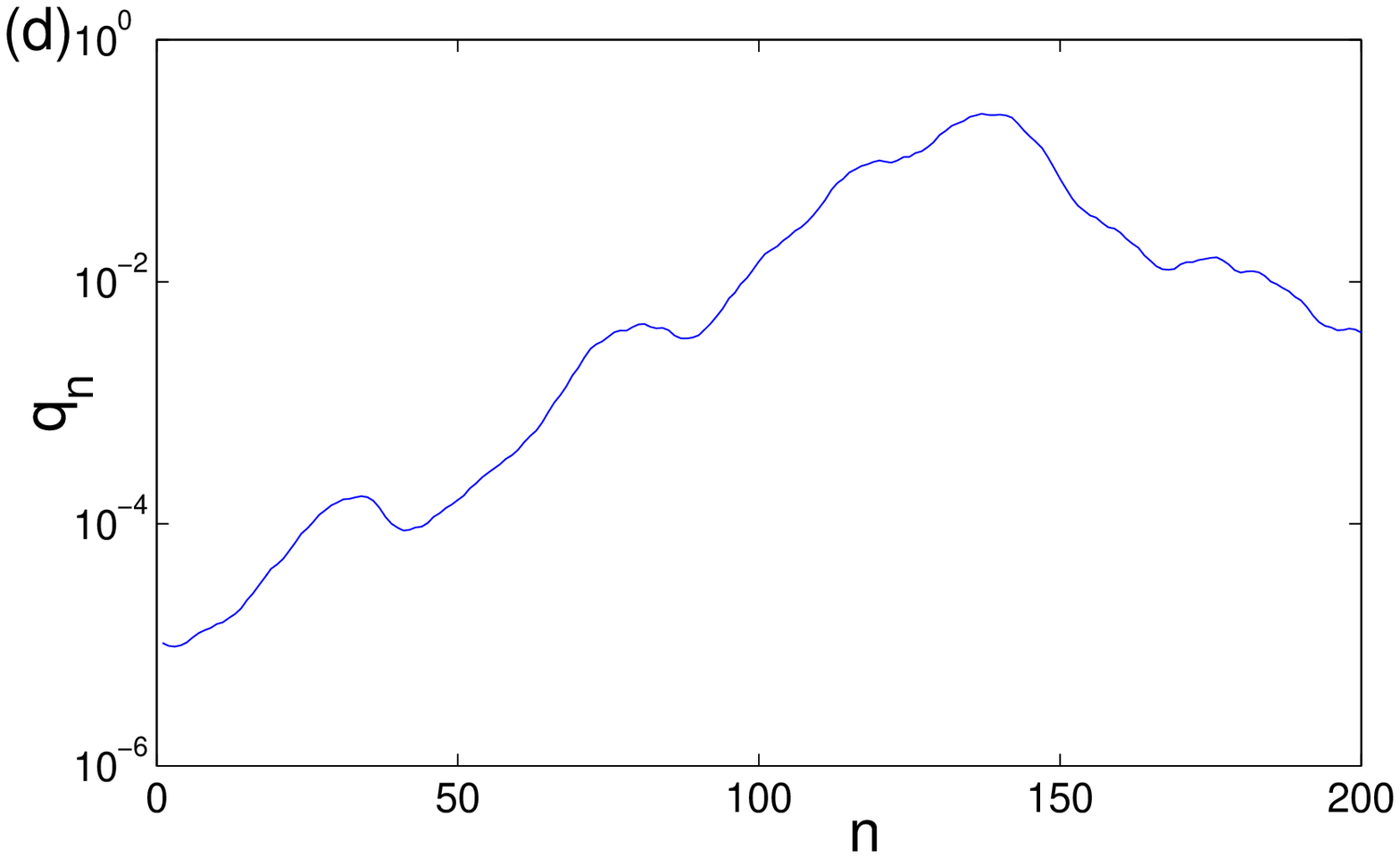}
	} \\
	\mbox{
	\includegraphics[width=6.4cm]{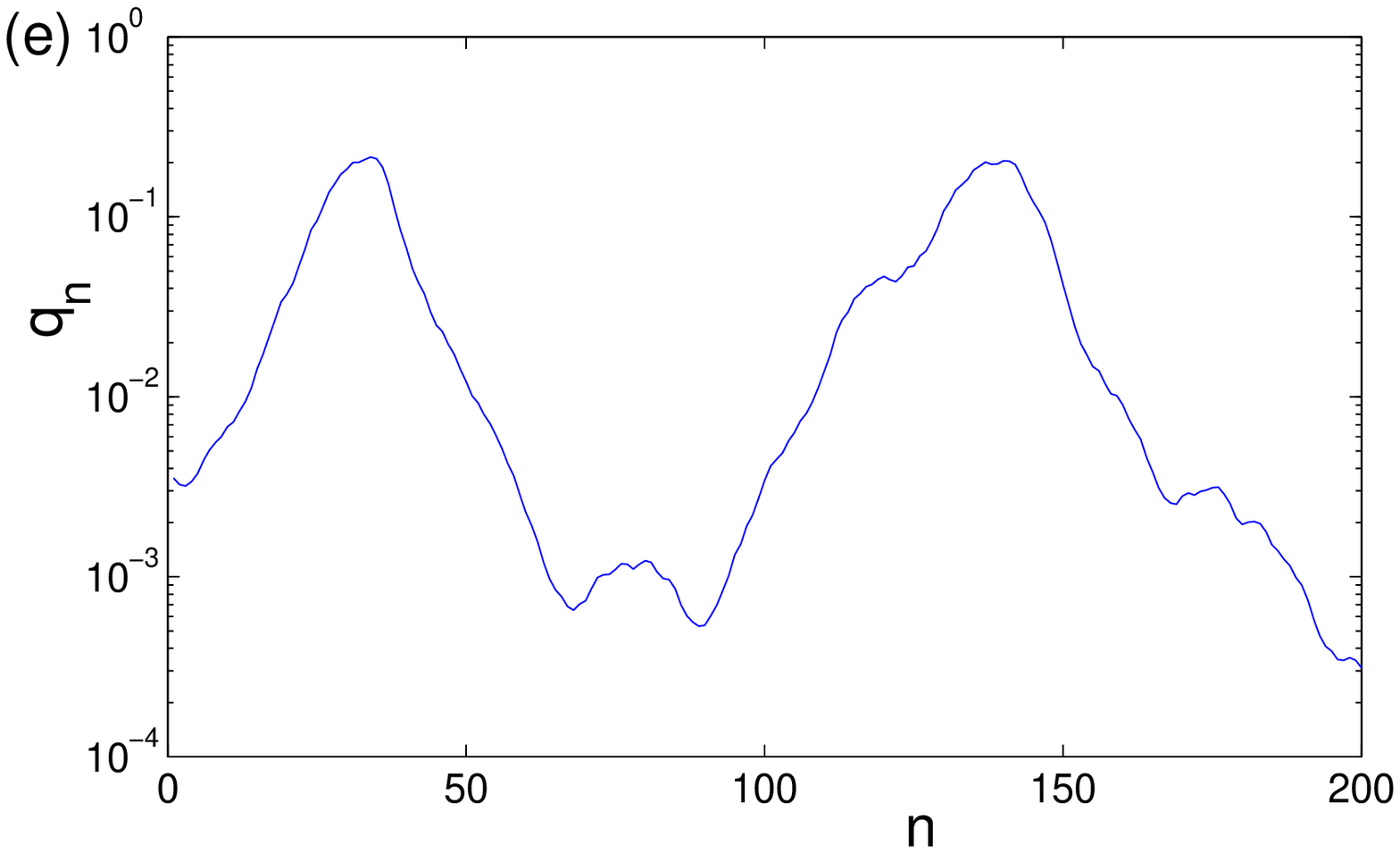}	
	\includegraphics[width=6.4cm]{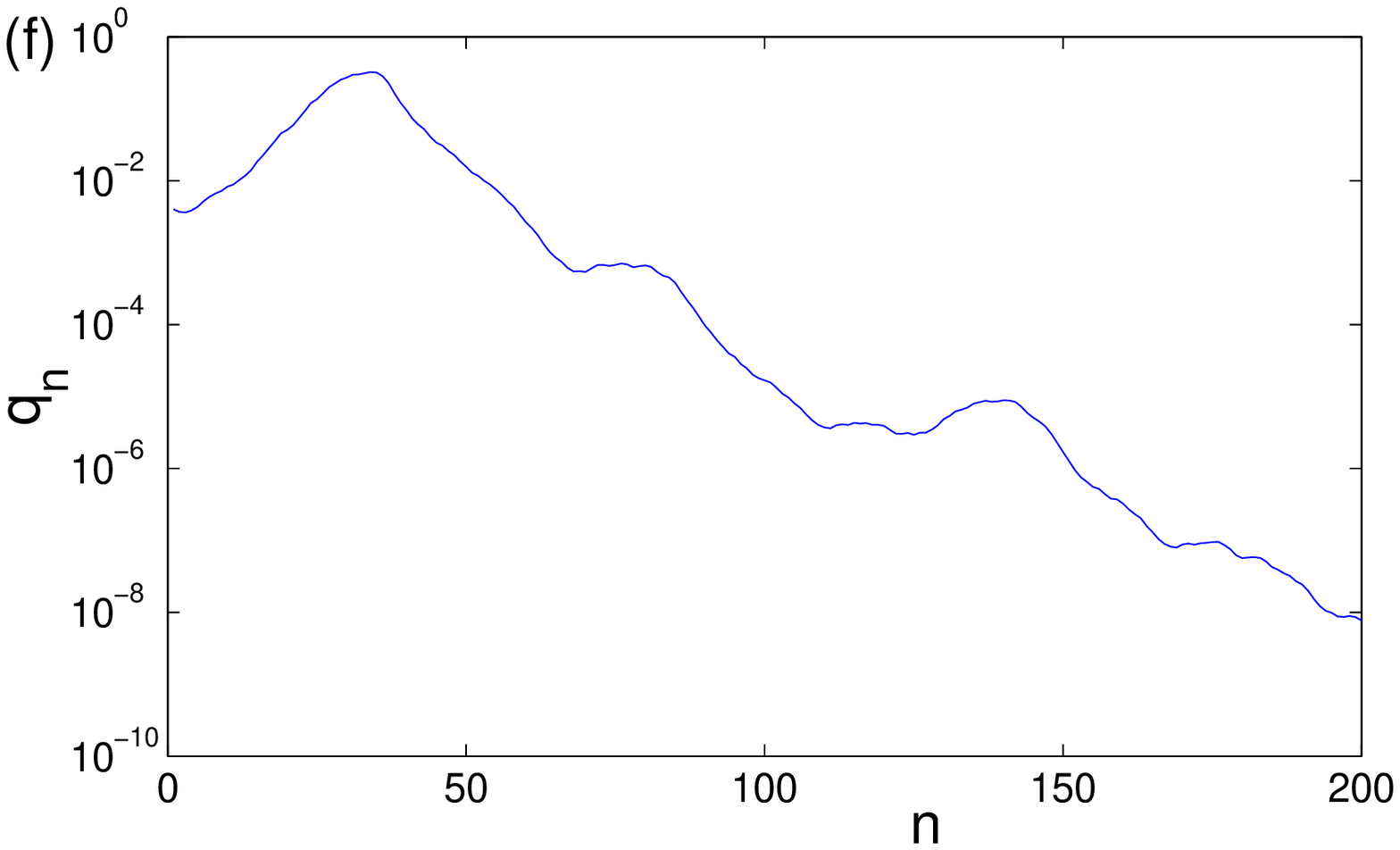}
	}
\caption{\small Synchronization analysis in a one dimensional chain of $N=200$ nonidentical Kuramoto phase oscillators according to Eq.~(\ref{Eq:diskr1}) and independent, identically, uniformly distributed random frequencies $\eta_n$ (values are shifted so that the mean $\bigl\langle\eta\bigr\rangle_\textnormal{System}=0$ is exactly zero).
(b) frequency values $-\eta_n$ corresponding to the potential of the discrete Hamiltonian (solid blue line) and a Gaussian filtering of width 2 (bold red line). There are several potential regions at which the ground state could be localized. (a) and (c) largest eigenvalues (solid blue lines) of the negative Hamiltonian $-\mathbf{H}$ in dependence on the heterogeneity $\sigma$, the second order perturbation approximation (Eq.~(\ref{Eq:PertResult003}), dashed red line), and the value $\sigma_{cr}$ for which the ground state becomes quasi degenerate (dashed black line). The quality of the approximation can be seen in double logarithmic scales in subfigure (c). (d)-(f) numerically determined groundstate eigenvectors in a semilogarithmic scale for (d) $\sigma=0.04$, (e) $\sigma=\sigma_{cr}=0.065135$ near the point of quasi degeneracy and (f) $\sigma=0.07$. Exponential localization at a potential well corresponds to concentric waves around this pacemaker region in the Kuramoto phase diffusion equations \cite{BlaToe05,Sakaguchi88}.}
\label{fig:RandCrossing}
\end{figure}

\begin{figure}[!htp]
	\center
	\mbox{
	\includegraphics[width=6.4cm]{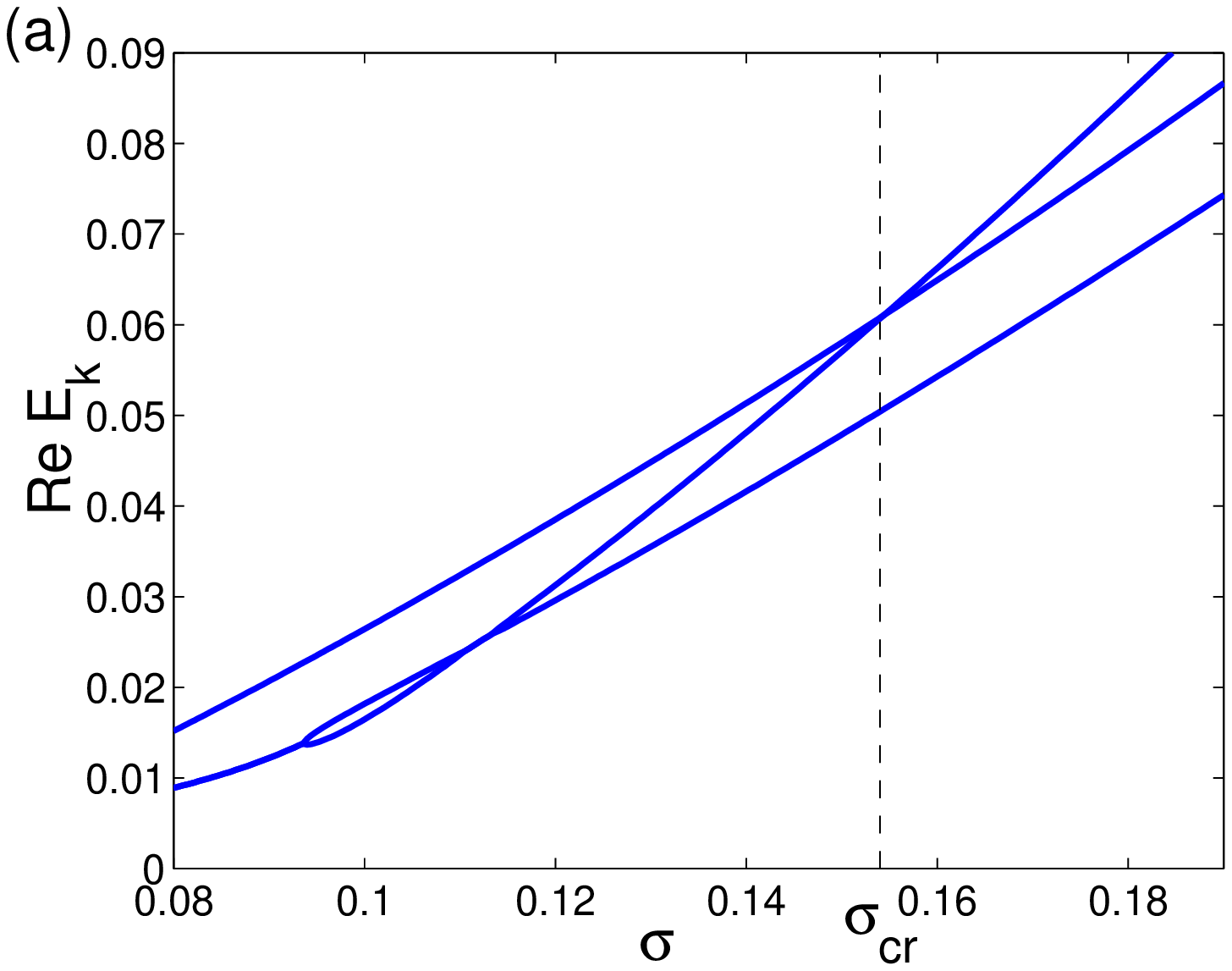} 
	\includegraphics[width=6.4cm]{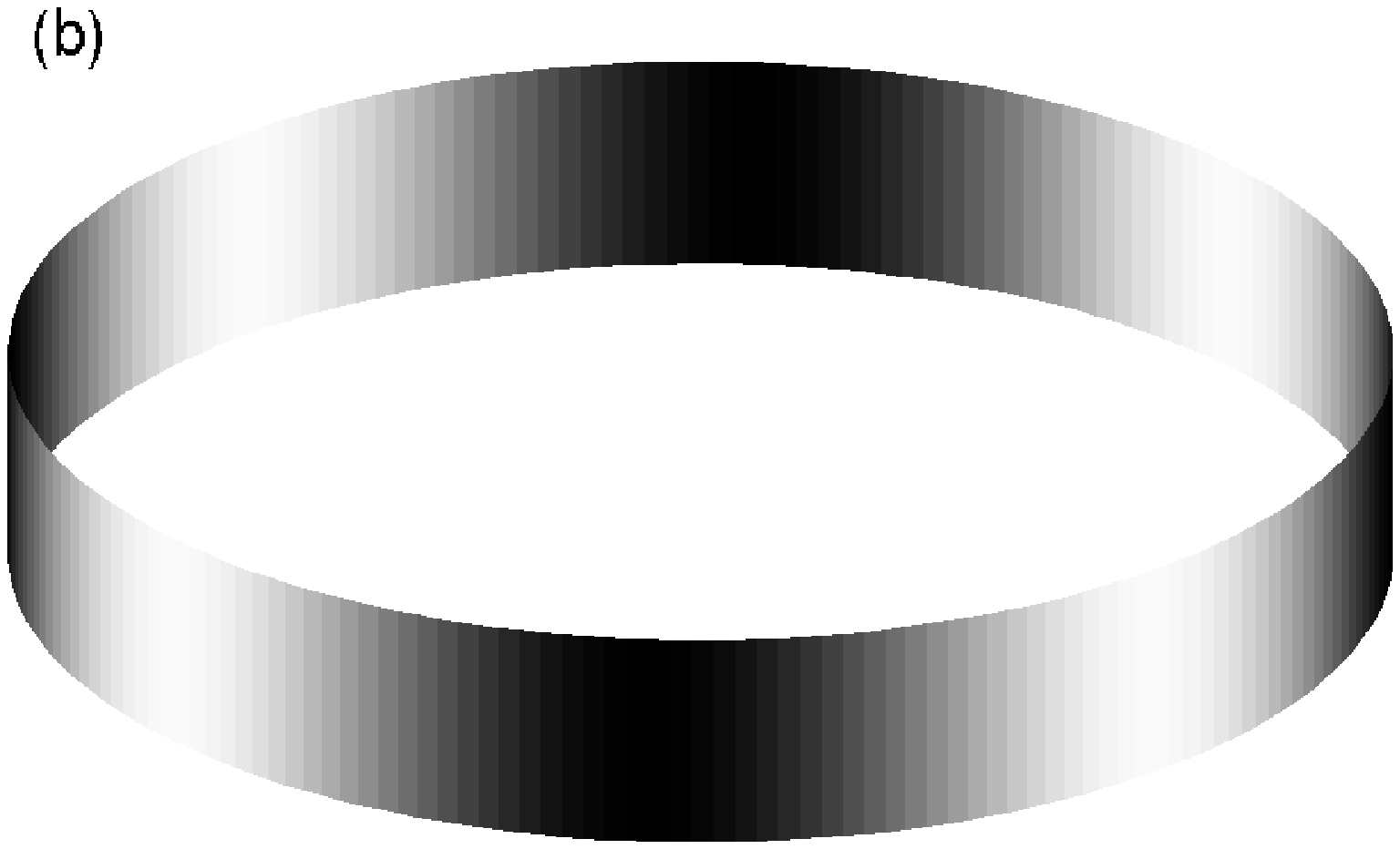}
	} \\
	\mbox{
	\includegraphics[width=6.4cm]{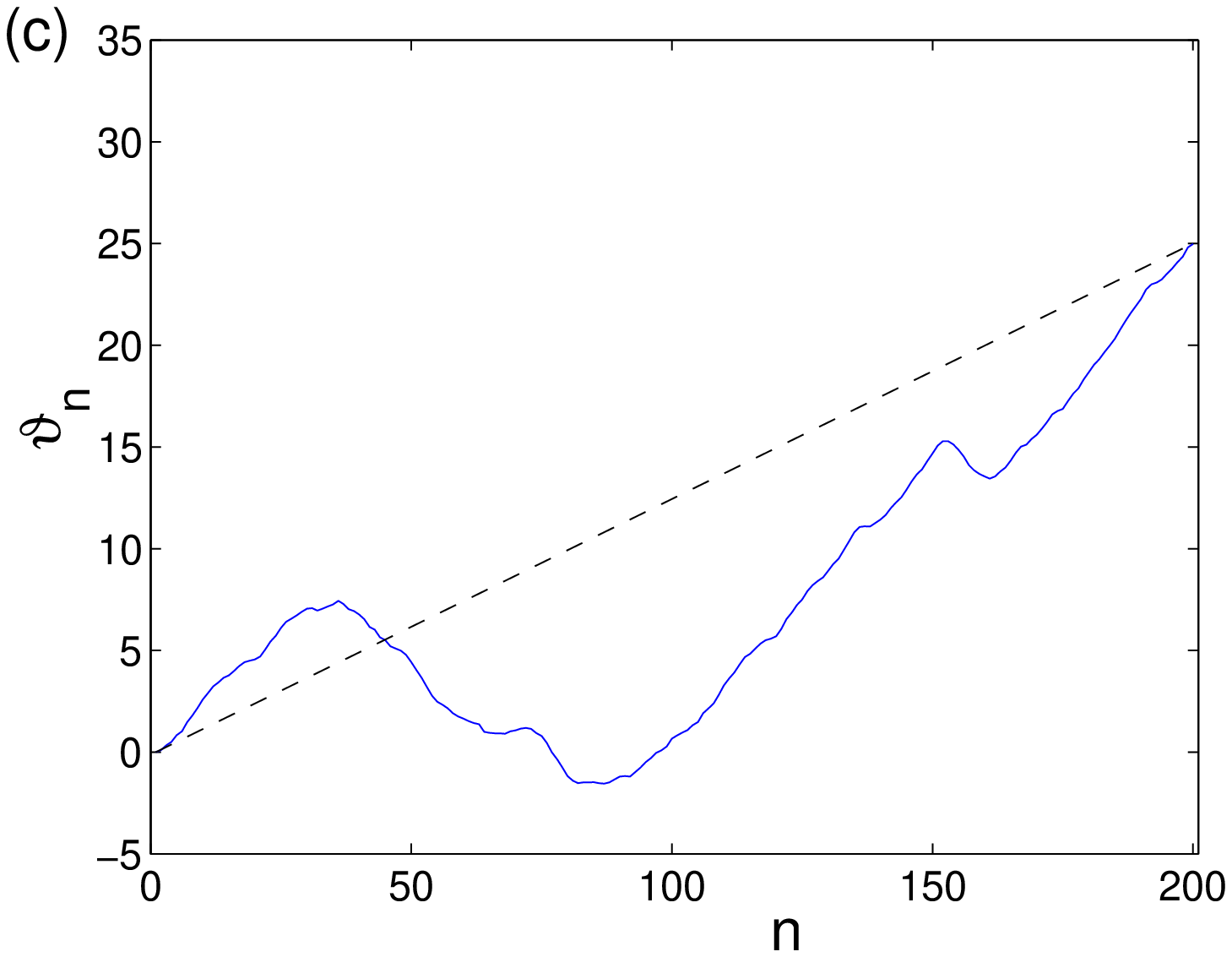}
	\includegraphics[width=6.4cm]{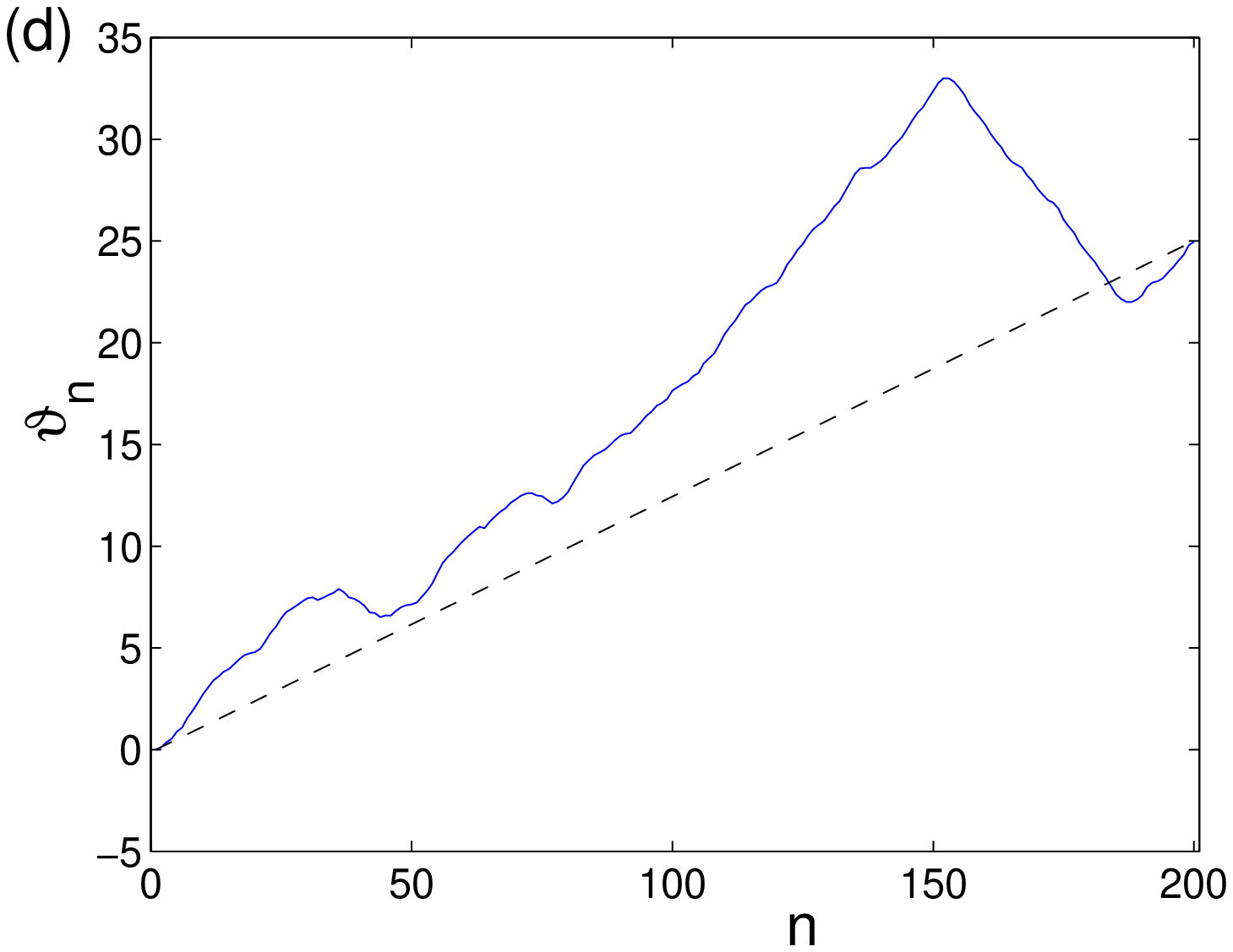}
	}
\caption{\small 
Synchronization analysis in a one dimensional chain of $N=200$ nonidentical Kuramoto phase oscillators with periodic boundary conditions and independent, identically distributed normal random frequencies $\eta_n$ (values are shifted so that the mean $\bigl\langle\eta\bigr\rangle_\textnormal{System}=0$ is exactly zero). The topological charge is $l=4$. (a) three largest eigenvalue real parts $\textnormal{Re}E_0>\textnormal{Re}E_1>\textnormal{Re}E_1$ of the difference operator in Eq.~(\ref{Eq:diskr1}) under variation of the heterogeneity $\sigma$. The second and third eigenvalue in this example are complex conjugated for $\sigma<0.09$, and the ground state becomes quasi degenerate for $\sigma_{cr}\approx 0.154$. (b) unperturbed ($\sigma=0$) rotating wave solution on a ring of oscillators, here as the sine of the phase (in grey levels) on the side of a cylinder for illustration. (c, d) stationary phase profiles of the corresponding discrete KPE (Eq.~(\ref{Eq:KPEs01}), blue lines) for $\sigma=0.15$ and $\sigma=0.16$ respectively, and the dashed black line is the unperturbed constant phase gradient.}
	\label{fig:RingCrossing}
\end{figure}

The approach to the synchronization problem based on Schr\"odinger perturbation theory has another, conceptual advantage. Upon the variation of a perturbation system parameter the eigenvalues of a Hamiltonian can become degenerate or quasi degenerate. The perturbation theory of finite order must fail before such a point. This effect is illustrated in Figure (\ref{fig:DoubleCrossing}), which investigates the difference equation 
\beq
\label{Eq:diskr1} 
E_k~p_n = \sigma\eta_n p_n + p_{n-1} - 2p_n + p_{n+1}	\quad,
\eeq
an approximation of Eq.~(\ref{Eq:Schroed01}) with $\gamma=1$ on a one-dimensional lattice and with open boundary conditions.
Two pacemaker regions of different size and natural frequency are competing as wave centers. For low natural frequencies the larger and slower  pacemaker region is dominating. By increasing both frequencies by a common factor $\sigma$, the smaller and faster region gains advantage. Two centers of waves can coexist in a small neighborhood around a critical value $\sigma_{cr}$. 
Since the ground state of a one dimensional Schr\"odinger equation cannot be degenerate for a potential with finite square integral norm the largest and the second largest eigenvalue never coincide. 
The levels can come exponentially close depending on the distance between the two potential wells. While the location of the dominating wave center shifts quickly upon variation of $\sigma$ the transient time until dominance is established scales as $|E_0-E_1|^{-1}$. In practice, since close to the critical parameter value only transient behavior can be observed, one cannot say how far the boundary of the concentric waves will shift in either direction.
The same transition can occur in a heterogeneous system with random frequencies as illustrated in Figures (Figs.~\ref{fig:RandCrossing},\ref{fig:RingCrossing}). The question where the pacemaker region of a heterogeneous oscillatory medium  is located given the local frequencies $\omega(\mathbf{r})$ cannot easily be answered without fully 
solving equation Eq.~(\ref{Eq:SchroedGauge01}) numerically.

\section*{VI. Discussion} 

In our study we have investigated the nonlinear Kuramoto Phase-Diffusion Equation (KPDE) in synchronization. 
We have applied perturbation theory to calculate the synchronized state in a heterogeneous oscillatory medium. To our knowledge we have presented the first explicit analytical results regarding the oscillation frequency and the phase profile in such a system.
Further we have identified different scaling relations depending on the system size and dimension and the frequency correlation length.
We have shown that the perturbation approach can straightforwardly be applied to simple topologies.
\\ \\ 
The first two terms of the perturbation series Eqs.~(\ref{Eq:OhmAnsatz02}-\ref{Eq:PertResult04}) 
are intuitively quite meaningful. If a medium with random frequencies synchronizes to a common synchronization frequency $\Omega$ then one expects $\Omega$ to be close to the mean frequency in the system. But this is exactly the first order perturbation term. Any deviation from the mean frequency is due to the nonlinearity $\gamma$ which appears as a factor only for higher order perturbation terms.    
\\ \\ 
Solutions of the Schr\"odinger Equation in disordered media are known to exhibit localization transitions \cite{Anderson58,KuzNies04}, depending on the system dimension and the strength of the disorder. Given equation Eq.~(\ref{Eq:Schroed01}) and the properties of the disordered potential, all the results from condensed matter physics \cite{KuzNies04,HalpLax66,HofstEtAl94} dealing with the localized states in the impurity band, and in particular the ground state, can in principle be applied. One of the results is that in one and two dimensions all states are localized. It is straightforward to show that in the limit of infinite system size
perturbation theory does, in fact, not work for localized states (cf. one dimensional delta potential). 
However, a perturbation ansatz is justified for states with a localization length larger than the system size. We have shown in the examples that the perturbation terms scale and diverge with the system size in one and two dimensional media. The result Eq.~(\ref{Eq:Scaling02}) suggests that $d=2$ is the critical dimension for the scaling of the synchronization frequency with the system size and the frequency correlation length. In one and two dimensions our perturbation theory only gives good quantitative predictions for finite systems with $\gamma\sigma L \ll 1$, where $L$ is the length of the system. In higher dimensions the synchronization frequency exists in the thermodynamical limit of $L\to\infty$ but it scales with the correlation length $\lambda$ of the frequencies as $\lambda^{2-d}$.  
\\ \\ 
The two presented perturbation approaches are equivalent in the sense that they lead to the same expressions for the perturbation terms, but when applied to a specific realization of frequencies it can be of advantage to choose one approach over the other. By reducing the nonlinear KPDE in synchronization to an eigenvalue problem one can find the ground state energy and the corresponding phase profile in the discretized system to arbitrary order precision using linear algebra methods. Special attention must be given, if the phase profile spans phase differences over several decades, i.e. when the system size is large compared to the wave length. Then the exponentially localized ground state must be computed to high precision even in the regions where it is several hundred orders of magnitudes smaller than at the localization point. For small nonlinearity $\gamma$ this second approach has no advantage over using the perturbation method Eqs.~(\ref{Eq:Perterm00}, \ref{Eq:Perterm01}, \ref{Eq:Perterm02}) on the KPDE directly. In particular, the nonlinear Cole-Hopf transformation (\ref{Eq:ColeHopf01}) introduces additional numerical errors.
\\ \\
We thank Professor A. S. Mikhailov for valuable discussions and Prof. J. Kurths for the support of this work.
This work was also supported by the DFG through the SFB555 and the Volkswagen Foundation.

\newpage

\clearpage

\end{document}